\documentclass[preprints,article,accept,pdftex,moreauthors]{Definitions/mdpi} 
\firstpage{1} 
\pubvolume{1}
\issuenum{1}
\articlenumber{0}
\pubyear{2022}
\copyrightyear{2022}
\datereceived{} 
\dateaccepted{} 
\datepublished{} 
\hreflink{https://doi.org/} 

\usepackage{amsmath,amsfonts,xspace, siunitx}

\newcommand{\dd}[1]{\ensuremath{\mathrm{d}#1}\xspace}
\newcommand{\dtRTR}{\ensuremath{\Delta t_\textrm{RTR}}}
\newcommand{\dtRTRdot}{\ensuremath{\Delta \dot{t}_\textrm{RTR}}}
\newcommand{\dtauRTR}{\ensuremath{\Delta \tau_\textrm{RTR}}}
\newcommand{\epstime}{\ensuremath{\epsilon_\textrm{time}}}
\newcommand{\xT}{\ensuremath{\textrm{T}}}
\newcommand{\xR}{\ensuremath{\textrm{R}}}
\newcommand{\xA}{\ensuremath{\textrm{A}}}
\newcommand{\xB}{\ensuremath{\textrm{B}}}
\newcommand{\xM}{\ensuremath{\textrm{M}}}

\usepackage{array}

\let\oldequation\equation
\let\oldendequation\endequation

\renewenvironment{equation}
  {\linenomathNonumbers\oldequation}
  {\oldendequation\endlinenomath}
  
\let\oldalign\align
\let\oldendalign\endalign

\renewenvironment{align}
  {\linenomathNonumbers\oldalign}
  {\oldendalign\endlinenomath}  

\Title{Comparing GRACE-FO KBR and LRI Ranging Data with Focus on Carrier Frequency Variations}

\TitleCitation{Comparing GRACE-FO KBR and LRI ranging data with focus on carrier frequency variations}

\Author{Vitali Müller $^{1}$*\orcidA{}, Markus Hauk$^{1,2,3}$, Malte Misfeldt $^{1}$\orcidC{}, Laura Müller $^{1}$, Henry Wegener $^{1}$, Yihao Yan $^{1}$, and Gerhard Heinzel $^{1}$*}

\AuthorNames{Müller, V.; Hauk, M.; Misfeldt, M.; Müller, L.;Wegener, H.;Yan, Y.; Heinzel, G.;}

\AuthorCitation{Müller, V.; Hauk, M.; Misfeldt, M.; Müller, L.;Wegener, H.;Yan, Y.; Heinzel, G.;}

\address{%
$^{1}$ \quad Max-Planck-Institut f\"ur Gravitationsphysik (Albert-Einstein-Institut) and Institut f\"ur Gravitationsphysik, Leibniz Universit\"at Hannover, Callinstrasse 38, D-30167 Hannover, Germany; \\
$^{2}$ \quad Deutsches Zentrum für Luft- und Raumfahrt (DLR), Institut für Satellitengeodäsie und Inertialsensorik,  Callinstr. 30b, 30167 Hannover, Germany; \\
$^{3}$ \quad Helmholtz Center Potsdam - GFZ German Research Centre for Geosciences, Department 1: Geodesy, Potsdam, Germany\\
}

\corres{Correspondence: vitali.mueller@aei.mpg.de (V.M.) and gerhard.heinzel@aei.mpg.de (G.H.)}

\abstract{The GRACE Follow-On satellite mission measures distance variations between the two satellites in order to derive monthly gravity field maps, indicating mass variability on Earth on a few 100\,km scale due to hydrology, seismology, climatology and others. This mission hosts two ranging instruments, a conventional microwave system based on K(a)-band ranging (KBR) and a novel laser ranging instrument (LRI), both relying on interferometric phase readout. In this paper we show how the phase measurements can be converted into range data using a time-dependent carrier frequency (or wavelength) that takes potential intraday variability in the microwave or laser frequency into account. Moreover, we analyze the KBR-LRI residuals and discuss which error and noise contributors limit the residuals at high and low Fourier frequencies. It turns out that the agreement between KBR and LRI biased range observations can be slightly improved by considering intraday carrier frequency variations in the processing. Although the effect is probably small enough to have little relevance for gravity field determination at the current precision level, the analysis is of relevance for detailed instrument characterization and potentially for future more precise missions. }


\begin{document}

\setcounter{section}{0} 

\section{Introduction}
The Gravity Recovery and Climate Experiment (GRACE) satellites have measured temporal variations in Earth's gravity field from 2002 until 2017. The successor mission GRACE Follow-On (GRACE-FO), launched in 2018, continues the valuable dataset of monthly gravity field maps \cite{landerer2020extending}, which are used for instance in the fields of climate research \citep{tapley2019contributions}, hydrology \cite{landerer2012accuracy} and seismology \cite{han2006crustal}. Prominent GRACE(-FO) results have quantified the ice mass loss in regions such as Greenland and Antarctic over the past two decades \cite{ciraci2020continuity} and allowed to attribute the effect of mass influx into the oceans in the observed global mean sea level rise \cite{chen2020global}. Some changes in the gravity field are likely caused by groundwater depletion \cite{CHEN2014130}. The GRACE(-FO) results lead to more than 2600 publications, and many of them are frequently cited in reports of the Intergovernmental Panel on Climate Change (IPCC).
 
The primary measurement that contains the gravity field information in a GRACE-like mission is the biased inter-satellite range, which was measured to micrometer precision in GRACE using a microwave K-band ranging system (KBR). GRACE-FO hosts two ranging instruments, the KBR and a novel and more precise laser ranging instrument (LRI, \cite{kornfeld2019grace, ghobadi2020grace}).  The ranging data of either KBR or LRI, or a combination of both, is processed in the course of gravity field retrieval together with various other measurements, such as attitude and orbit data as well as with measurements of non-gravitational accelerations from accelerometers. 

LRI and KBR measure the inter-satellite distance variations in GRACE-FO in parallel. Future missions such as the European-lead Next Generation Geodesy Mission (NGGM, \cite{nicklaus2020laser}) and the US-lead Mass Change Mission \cite{wiese2020nasa}, with counterpart GRACE-I(carus) \,\cite{flechtner2020realization} in Germany, are currently being studied or developed. Since these new missions will solely rely on laser-based ranging, it is important to understand what is currently limiting the ranging data. 

The GRACE-FO data discussed here is publicly available in terms of raw observations (level1a), in processed observations (level1b) and in terms of gravity field products (level2). The unique setup of two almost independent ranging systems allows us to study the differential instrument behavior and small instrument errors at frequency regions which are usually dominated by signal several orders of magnitude larger than the expected noise.

In this paper, we compare the LRI and KBR ranging data in the spectral domain, revisit the main error and noise contributors in both instruments and attempt to explain the current level of KBR-LRI residuals. We pay special attention to carrier frequency variations in both instruments and on the approach to account for them when converting the observed phase (level1a) to a range (level1b).

\section{Phase Observable}
\label{sec:PhaseObservable}
\begin{figure}
    \centering
    \includegraphics[width=14.0cm]{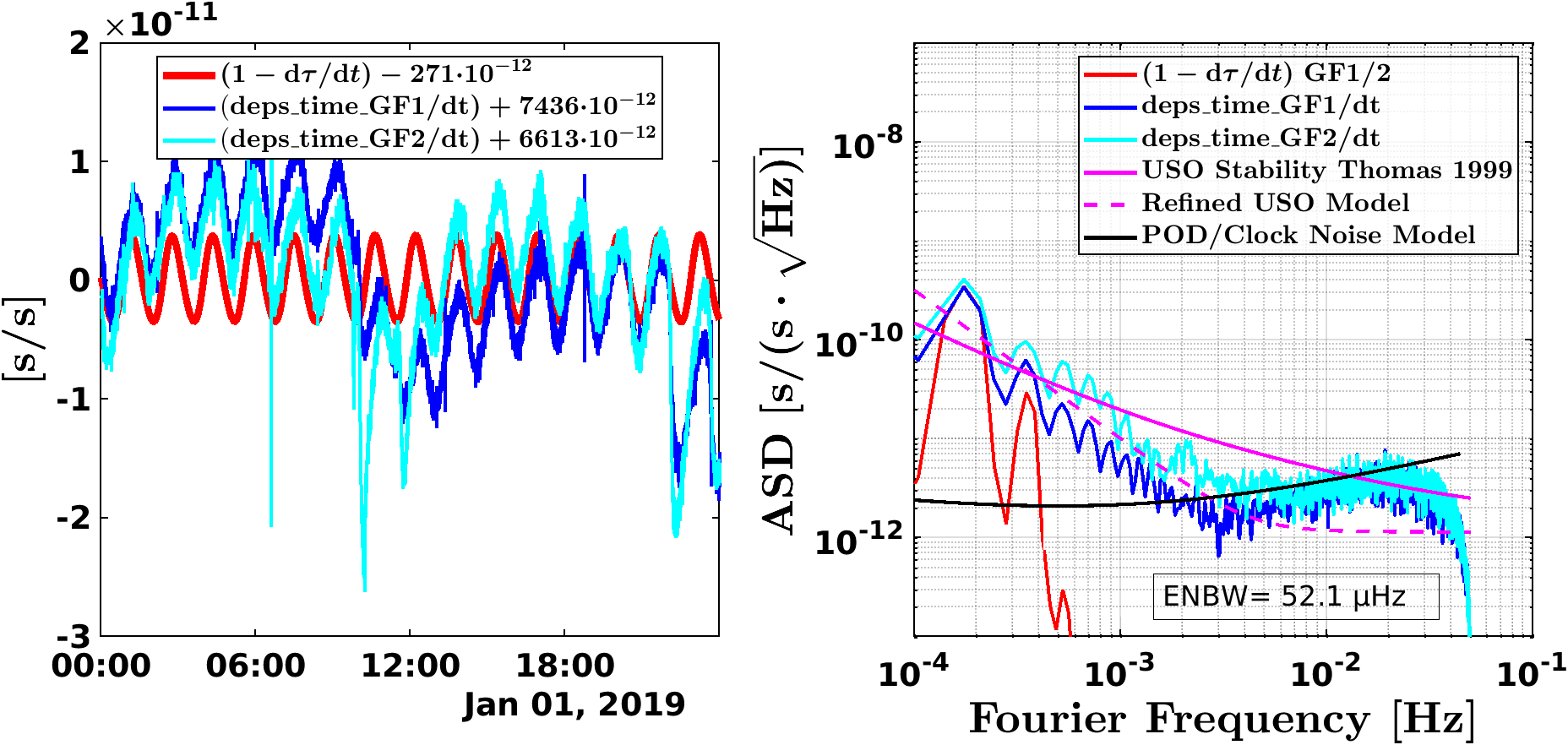}
\caption{(Left:) Typical time-domain plot of the proper time $\tau$ (red) determined from the satellite orbit product and the CLK1B \epstime (light and dark blue), both in terms of the first derivative. The proper time is clearly visible in \epstime. (Right): Amplitude spectral density (ASD) of the time-series shown on the left side together with an estimate of the USO stability based on \cite{thomas1999analysis}. The traces labelled refined USO model and POD/Clock noise model are described in sec.~\ref{sec:TimetagError}. See fig.~\ref{fig:AllanDeviation} for a representation as Allan deviations. The Equivalent Noise Bandwidth (ENBW) given here and in other spectra can be used to convert the spectral density units to  spectrum (rms) units \cite{Heinzel.2002}. }
    \label{fig:propertime}
\end{figure}

The phase of microwave or optical radiation can be described in a general-relativistic context as
\begin{align}
    \Phi = \nu_0 \cdot \tau_\mathrm{osc}(t) + q_0 = \nu_0 \cdot [\tau(t) + \tau_\mathrm{FV}(\tau(t))]  + q_0,
\end{align}
where $t$ is the coordinate time or GPS time, $\nu_0$ the nominal frequency, and $q_0$ is a constant that cannot be resolved in general using interferometry, because phase observations exhibit an integer-ambiguity. $\tau_\mathrm{osc}$ is a virtual oscillator time that is given by the sum of the proper time $\tau$ of the satellite or emitting device (e.g. laser) and a contributor $\tau_\mathrm{FV}$ describing frequency variations or, in other words, deviations of the oscillator time from proper time due to imperfections, noise, etc. The oscillator time for the KBR is realized inside the instrument processing unit (IPU) of the microwave ranging instrument (MWI) assembly and can be called the ultra-stable oscillator (USO) time $\tau^\textrm{USO}$ or the IPU receiver time. The difference between USO time and GPS time is the clock error, which is estimated during precise orbit determination (POD) together with the satellite's position and velocity state. The clock error is reported as time series in the CLK1B data product as a \textrm{eps\_\textrm{Time}} (\epstime), i.e.
\begin{align}
    \epstime = t-\tau^\textrm{USO}(t). \label{eq:EpsTimeDef}
\end{align}
\epstime~contains the proper time, clock variations and a frequency offset, as can be seen from fig.~\ref{fig:propertime}. The proper time for a satellite can easily be computed by numerically integrating \cite{burt2021demonstration}
\begin{align} \label{eq:dtau-1}
    1-\frac{\dd{\tau}}{\dd{t}} &= \frac{GM}{rc_0^2} - \frac{GM}{rc_0^2}\cdot J_2 \cdot \Bigl(\frac{a_e}{r}\Bigr)^2 \cdot \frac{3z^2-r^2}{2r^2} + \frac{v^2}{2c_0^2}+ \frac{\phi_0}{c_0^2},
\end{align}
where $c_0$ is speed of light, $GM$ is Earth's gravitational parameter, $a_e$ is Earth's mean readius, $J_2$ is Earths oblateness coefficient , $r$ is the distance between geocenter and satellite, $z$ is the z-component of the position vector, $v$ is the satellite velocity and $\phi_0$ is a potential constant.

The frequency offsets of the USOs w.r.t. its nominal frequency yield y-offsets in the left plot of fig.~\ref{fig:propertime}. As apparent from the legend, they have values of \SI{7.4}{ppb} on satellite~1 (GF-1) and \SI{6.6}{ppb} on satellite~2 (GF-2) on that particular day.
The oscillator time of the laser is not accessible and is usually expressed in terms of the laser frequency. 
The instantaneous frequency reads
\begin{align}
    \nu(\tau) := \frac{ \dd \Phi }{ \dd \tau }(\tau) = \nu_0 \cdot \left( 1 + \frac{\dd  \tau_\mathrm{FV} }{ \dd \tau } \right),
\end{align}
where $\nu_0 \cdot \dd  \tau_\mathrm{FV} / \dd \tau $ describes the frequency variations or noise with respect to a nominal frequency $\nu_0$.

The phase $\Phi$ is invariant under Lorentz transformations \cite{shiozawa2004classical}, which means it does not change its value when transformed from the local system of the satellite to the geocentric celestial reference system (GCRS), e.g. 
\begin{align}
    \Phi(\tau_\xA, \vec r_\xA) = \Phi(t_\xA, \vec r_\xA^{~\prime}),
\end{align}
where $(\tau_\xA, \vec r_\xA)$ is the four-valued event in the local satellite system, while $(t_\xA, \vec r_\xA^{~\prime})$ denotes the same event in the GCRS.

The frequency is not relativistically invariant, since it appears in the GCRS as
\begin{align}
    \nu(t) := \frac{ \dd \Phi }{ \dd \tau } (\tau(t)) \cdot \frac{\dd \tau}{\dd t} (t) = \nu_0 \dot{\tau}(t) \cdot \left( 1 + \frac{\dd  \tau_\mathrm{FV} }{ \dd \tau } (\tau(t)) \right), \label{eq:NuInGps}
\end{align}
where the short form for the derivative $\dot{\tau} = \dd \tau / \dd t$ was used. We note that even if a perfect laser system would produce a constant freqeuncy in the satellite frame, the frequency is varying in other systems like the geocentric one used for gravity field determination.

The optical phase is the time-integral of the optical frequency, i.e.~
\begin{align}
  \Phi = \int_{\tau_0}^{\tau}    \nu(\tau^\prime) ~ \dd \tau^\prime = \int_{t(\tau_0)}^{t(\tau)}    \nu(t^\prime)~\dd t^\prime. \label{eq:PhaseFrequencyRelation}
\end{align}

\section{LRI Ranging Phase}
\label{sec:RangingPhaseLRI}
For the LRI, the ranging phase $\varphi_\textrm{TWR}$ is called a two-way ranging (TWR) quantity in the GRACE-FO project and it is determined from the phase difference between transponder $\varphi_\xT$ and reference $\varphi_\xR$ satellite \cite{GFO_Level1_Handbook}. The phase measurements on both satellites are performed with phase meters. The transponder satellite uses a phase-locked loop to lock its laser to the incoming light using a high gain and high bandwidth control loop \cite{abich2019orbit}. This means the transponder phase is a rather trivial measurement given by a phase ramp with constant slope \cite[eq.~(2.207)]{mueller2017design}
\begin{align}
 \varphi_\xT = \SI{10}{MHz} \cdot \tau^\textrm{USO}_\xT, \quad [\varphi_\xT]=\textrm{cycles}. \label{eq:TransponderPhaseBasic}
\end{align}
On the reference satellite, the laser is stabilized using an optical cavity and the phase meter output $\varphi_\xR$ contains both the ranging information and the transponder phase ramp. Since the same phase ramp is present on both satellites, it cancels out in the differential measurement $\varphi_\textrm{TWR}$ and one can write the TWR phase as \cite{Yan2020}
\begin{align}
      \varphi_\textrm{TWR} = \varphi_\xT (t - \Delta t_\textrm{TR}) - \varphi_\xR(t) = \Phi_\xR(t) - \Phi_\xR(t -\dtRTR) + q_\xR , \label{eq:varphiTWR}
\end{align}
where $\Phi_\xR$ is the optical phase of the reference laser in units of cycles and $\dtRTR$ is the round-trip light propagation time from reference to transponder and back to the reference satellite. The round-trip propagation time $\dtRTR$ contains the ranging information, while the propagation time from transponder to reference satellite $\Delta t_\textrm{TR}$ can be derived from GPS orbit data and is needed to account for the propagation of the phase value from transponder to reference. The constant $q_\xR$ arises from the inability of interferometers to measure the absolute phase or distance. 

Using the definition of the optical phase from eq.~(\ref{eq:PhaseFrequencyRelation}), one obtains
\begin{align}
    \varphi_\textrm{TWR}&= \int^{\tau}_{\tau -\dtauRTR} \nu_\xR(\tau^\prime) ~ \dd \tau^\prime + q_\xR \\
    &=\nu_\xR \cdot [ \tau-\tau+\dtauRTR(\tau) + \tau_\mathrm{FV}(\tau )-\tau_\mathrm{FV}(\tau -\dtauRTR )  ] + q_\xR \\
    &\approx \nu_\xR \cdot \dtauRTR + \nu_\xR \cdot \frac{\dd \tau_\mathrm{FV}}{\dd \tau} \cdot \dtauRTR + q_\xR \\
    &= \nu_\xR(\tau) \cdot \dtauRTR(\tau)  + q_\xR  = \nu_\xR(t) \cdot \dtRTR(t) + q_\xR \label{eqPhir1a},
\end{align}
where the last line shows the phase observable in terms of proper or local quantities on the satellite ($\nu(\tau) \cdot \dtauRTR(\tau)$) and
in terms of GCRS quantities ($\nu(t) \cdot \dtRTR(t)$). We use $\varphi$ to indicate phase quantities that have a slope in the radio-frequency domain (kHz or MHz), i.e.~which are tracked and recorded on GRACE-FO, while the phase quantities $\Phi$ have phase slopes in the THz or GHz domain, which cannot be recorded.

The LRI level1a (LRI1A) data of GRACE-FO contains the raw phases of reference $\varphi_\xR$ and transponder $\varphi_\xT$ satellite, while the LRI1B product contains the raw biased range $\rho_\textrm{TWR}$ and the light-time correction  $\rho_\textrm{LTC}$. Adding the raw range and the LTC yields the corrected biased range 
\begin{align}
  \rho_\textrm{LRI} &= \rho_\textrm{TWR}( \varphi_\textrm{TWR} ) + \rho_\textrm{LTC,TWR} + \rho_\textrm{TTL} + \rho_\textrm{media} \\
  &\approx \rho_\textrm{TWR}( \varphi_\textrm{TWR} ) + \rho_\textrm{LTC,TWR} =  | \vec r_\xA(t) - \vec r_\xB(t) | + \mathrm{bias}
\end{align}
which corresponds to the biased instantaneous Euclidean distance between satellites in the GCRS and that serves as input for the gravity field recovery. The effects from tilt-to-length coupling $\rho_\textrm{TTL}$ and from the light propagation in a medium $\rho_\textrm{media}$ are negligible for LRI, as will be shown in sec.~\ref{sec:InstrumentNoises}.
 The light-time correction is computed as
\begin{align}
    \rho_\textrm{LTC,TWR} := | \vec r_\xA^\textrm{~GPS}(t) - \vec r_\xB^\textrm{~GPS}(t) |  - \frac{c_0 \dtRTR^\textrm{GPS}(t)}{2}
\end{align}
from GPS orbit data of the satellites and using formulas for the light time propagation time ($\dtRTR^\textrm{GPS}$) that depend on the orbit data  \cite{Yan2020}. The magnitude of $\rho_\textrm{LTC,TWR}$ is below a millimeter  such that the GPS accuracy of a few millimeter over the inter-satellite baseline of approx. \SI{200}{km} is sufficient in order to not limit the data quality of $\rho_\textrm{LRI}$.

The derivation of the range $\rho$ from the (round-trip) phase $\varphi$ is trivial, if one assumes a constant frequency $\nu$
\begin{align}
    \rho(t) = \frac{\lambda}{2} \cdot \varphi(t) = c_0  \cdot \frac{ \varphi(t) }{2\nu}. \label{eq:RhoNaive}
\end{align}
The wavelength $\lambda$ or frequency $\nu$ serves as a conversion factor from phase with units of cycles to a half round-trip range in units of meters. The currently available (v04) level1b data products of LRI and KBR assume a day-wise constant frequency such that the equation is applicable and used to convert data from individual days. 

However, such an approach has the disadvantage that the range data might have some offsets and slope changes at day bounds due to the daily changes of the frequency. Moreover, we anticipate that intraday frequency variations $\nu(t)$ are a noticeable error term in both laser and microwave interferometers. Thus, we investigated methods to convert the full round-trip phase  $ \varphi_\textrm{TWR}$  to a half round-trip range $\rho_\textrm{TWR}$ when frequency variations are perfectly known. This means we are searching for functions $\rho_\textrm{TWR}( \varphi_\textrm{TWR} )$ that, for the LRI, yield
\begin{align}
     \rho_\textrm{TWR}( \varphi_\textrm{TWR} )= c_0 \frac{\dtRTR(t)}{2} + \textrm{const}. \label{eqAim}
\end{align}
Once the error-free equations are understood, we address errors in the knowledge of $\nu$ in sec.~\ref{sec:ScaleError}.

\subsection{Formula 1: Phase-Frequency Ratio}
\label{sec:PhaseFreqRatio}
The direct naive approach using eq.~(\ref{eq:RhoNaive}) with a time-dependent frequency and approximation in eq.~(\ref{eqPhir1a}) yields
\begin{align}
    \rho_\textrm{TWR,1}(t) := c_0 \cdot \frac{ \varphi_\textrm{TWR}(t)}{2\nu_\xR(t)} \approx c_0 \frac{\dtRTR(t)}{2} + c_0\frac{q_\xR}{2\nu_\xR(t)}. \label{eq:rhoTWR1}
\end{align}
The divisor is the apparent frequency in the GCRS $\nu(t)$, since $\nu(\tau)$ would result in a contracted range $c_0 \dtauRTR$.
The naive formula $\rho_\textrm{TWR,1}$ yields the correct result of eq.~(\ref{eqAim}), if $q_\xR=0$, i.e.~if the phase measurement $\varphi_\textrm{RTR}$ is an absolute measurement without integer ambiguity due to indistinguishable cycles. In actual interferometry, the phase time-series $\varphi_\textrm{RTR}(t)$ exhibits an unknown arbitrary bias and it is debiased in order to avoid large numerical values, loss of numerical precision due to floating-point arithmetic or overflow in finite-sized registers of a computer. We assume without loss of generality that phase time-series start at 0\,cycles at an initial epoch, i.e. $\varphi_\textrm{RTR}(t=0) = 0$ in a stretch of data that is being converted from phase to range. This implies  that
\begin{align}
q_\xR &= -\nu_\xR(t=0) \cdot \dtRTR(t=0) \\
    \rho_\textrm{TWR,1}(t) &\approx c_0 \frac{\dtRTR(t)}{2} - c_0 \dtRTR(0) \cdot \frac{  \nu_\xR(t=0) }{2\nu_\xR(t)} \label{eq:Rho1}
\end{align}
where the first summand on the right-hand side in line~(\ref{eq:Rho1}) is the desired ranging signal and the second term describes laser frequency variations that couple via the baseline length $c_0\dtRTR(0)/2 \approx \SI{220}{km} $. Since eq.~(\ref{eq:Rho1}) shows the deviation of $\rho_\textrm{TWR,1}$ from the correct result, we can directly define an improved formula as
\begin{align}
    \rho_\textrm{TWR,1corr}(t) &:= c_0 \cdot \frac{ \varphi_\textrm{TWR}(t)}{2\nu_\xR(t)} + c_0 \Delta t^\textrm{GPS}_\textrm{RTR}(0) \cdot \left( \frac{\nu_\xR(0)}{2\nu_\xR(t)} - \frac{1}{2} \right), \label{eq:Rho1Corr1} \\ 
    &\approx c_0 \cdot \frac{ \varphi_\textrm{TWR}(t)}{2\nu_\xR(t)} - \frac{c_0 \Delta t^\textrm{GPS}_\textrm{RTR}(0)}{2} \cdot \left( \frac{\nu_\xR(t)-\nu_\xR(0)}{\nu_\xR(0)} \right), \label{eq:Rho1Corr1approx} 
\end{align}
where the second correction term on the right-hand side in line~(\ref{eq:Rho1Corr1}) is usually much smaller than the first ranging term, as the expression in the bracket is close to zero due to the introduced constant $-1/2$. The round-trip propagation time at the initial epoch received the superscript GPS ($c_0 \dtauRTR^\mathrm{GPS}(0)$) in order to indicate that the value can be derived from GPS orbit data and algorithms to compute the light propagation time \cite{Yan2020}. The millimeter to centimeter precision of GPS is sufficient to derive the absolute distance $c_0 \Delta t_\textrm{RTR}^\mathrm{GPS}(0) \approx \SI{440}{km}$ without spoiling the high precision from LRI in $\rho_\textrm{TWR,1corr}$, since the correction term is small due to $\nu_\xR(0)/\nu_\xR(t) \approx 1$. The approximation in line~(\ref{eq:Rho1Corr1approx}) was based on $1/(1+x) \approx 1-x$, with $x=\nu_\xR(t)/\nu_\xR(0)-1$, and it illustrates more clearly that the correction term is the product of inter-satellite distance and fractional frequency deviations.


\subsection{Formula 2: Integral of Differentiated Phase}
An alternative way to compute the half round-trip range from the round-trip ranging phase $\varphi_\textrm{TWR}$ has been suggested as \cite[eq.~(60)]{Yan2020}
\begin{align}
    \rho_\textrm{TWR,2} \approx c_0 \int^{t}_0  \frac{ \dd \varphi_\textrm{TWR}(t^\prime) / \dd t^\prime }{2 \nu_\xR(t^\prime)} ~\dd t^\prime \label{eq:Rho2Wrong}
\end{align}
though it is easy to see that the equation is not exact, since
\begin{align}
     \frac{ \dd \varphi_\textrm{TWR} }{\dd t}  &= \frac{\dd}{\dd t} \int_{t-\Delta t}^t \nu_\xR(t^\prime) ~\dd t^\prime  =  \nu_\xR(t) - \nu_\xR(t -\Delta t) \cdot (1 - \dd \Delta t / \dd t )  \\
     &= \nu_\xR(t -\Delta t) \cdot \dd \Delta t / \dd t  + \nu_\xR(t) - \nu_\xR(t -\Delta t)  \label{eq:Rho2b} \\
     &\approx \nu_\xR(t) \cdot \dd \Delta t / \dd t + (1 - \dd \Delta t / \dd t ) \cdot \dot{\nu}_\xR(t) \cdot  \Delta t  \label{eq:dPhiApprox}
\end{align}
The first term in the line (\ref{eq:Rho2b}) contains the ranging information that is multiplied with the optical frequency at the emission event and not at the reception event as the divisor in eq.~(\ref{eq:Rho2Wrong}) suggests. Moreover, the second and third terms indicate needed correction terms that are missing in eq.~(\ref{eq:Rho2Wrong}). Thus, the exact formula to convert the round-trip phase $\varphi_\textrm{TWR}$ to a half round-trip range reads
\begin{align}
    \rho_\textrm{TWR,2exact}(t) &:=\frac{c_0}{2} \int^{t}_0  \frac{ \dd \varphi_\textrm{TWR}(t^\prime) / \dd t^\prime }{ \nu_\xR \left(t^\prime - \dtRTR(t^\prime)\right)  } - \left( \frac{\nu_\xR(t^\prime) }{\nu_\xR \left(t^\prime - \dtRTR(t^\prime)\right)} - 1 \right) ~\dd t^\prime \label{eq:Rho2Corr} \\
    & = \frac{c_0 \dtRTR(t)}{2} - \frac{c_0 \dtRTR(0)}{2}\ . 
\end{align}
The term 
\begin{align}
    &\frac{c_0}{2}\int_0^t \frac{\nu_\xR(t^\prime)}{\nu_\xR(t^\prime-\Delta t_\mathrm{RTR}(t^\prime))} - 1\,\dd t^\prime \nonumber \\
    &\quad \approx \frac{c_0}{2}\int_0^t \frac{\dot{\nu}_\xR(t^\prime)}{\nu_\xR(t^\prime)} \cdot \Delta t_\mathrm{RTR}(t^\prime)\,\dd t^\prime \approx \frac{c_0}{2} \langle \Delta t_\mathrm{RTR} \rangle \cdot \frac{\log(\nu_\xR(t))}{\log(\nu_\xR(0))} \approx \frac{c_0}{2} \langle \Delta t_\mathrm{RTR} \rangle \cdot \frac{\nu_\xR(t)-\nu_\xR(0)}{\nu_\xR(0)} 
\end{align}
approximately describes the range correction due to laser phase changes on the reference satellite while the laser light propagates forth and back between satellites with average propagation time $\langle \Delta t_\mathrm{RTR} \rangle$. The last approximation of this expression resembles the correction term shown in eq.~(\ref{eq:Rho1Corr1approx}).

A handy approximation of the exact form in eq.~(\ref{eq:Rho2Corr}) with all quantities being evaluated at the same time is given as
\begin{align}
    \rho_\textrm{TWR,2approx} = c_0 \int^{t}_0  \frac{ \dd \varphi_\textrm{TWR}(t^\prime) / \dd t^\prime }{2 \nu_\xR \left(t^\prime\right)  } - \left( 1 - \frac{ \dd \dtRTR^\textrm{GPS} }{\dd t} \right) \cdot  \frac{\dot{\nu}_\xR(t^\prime) }{2\nu_\xR \left(t^\prime\right)} \dtRTR^\textrm{GPS}(t^\prime)   ~\dd t^\prime, \label{eq:Rho2approx}
\end{align}
which is based on the approximation shown in eq.~(\ref{eq:dPhiApprox}). The light propagation time in the correction term received a superscript GPS to indicate that this quantity can be obtained from the orbit product, as in eq.~(\ref{eq:Rho1Corr1}).

\subsection{Accuracy of Approximations}
\begin{figure}
    \centering
    \includegraphics[width=12.5cm]{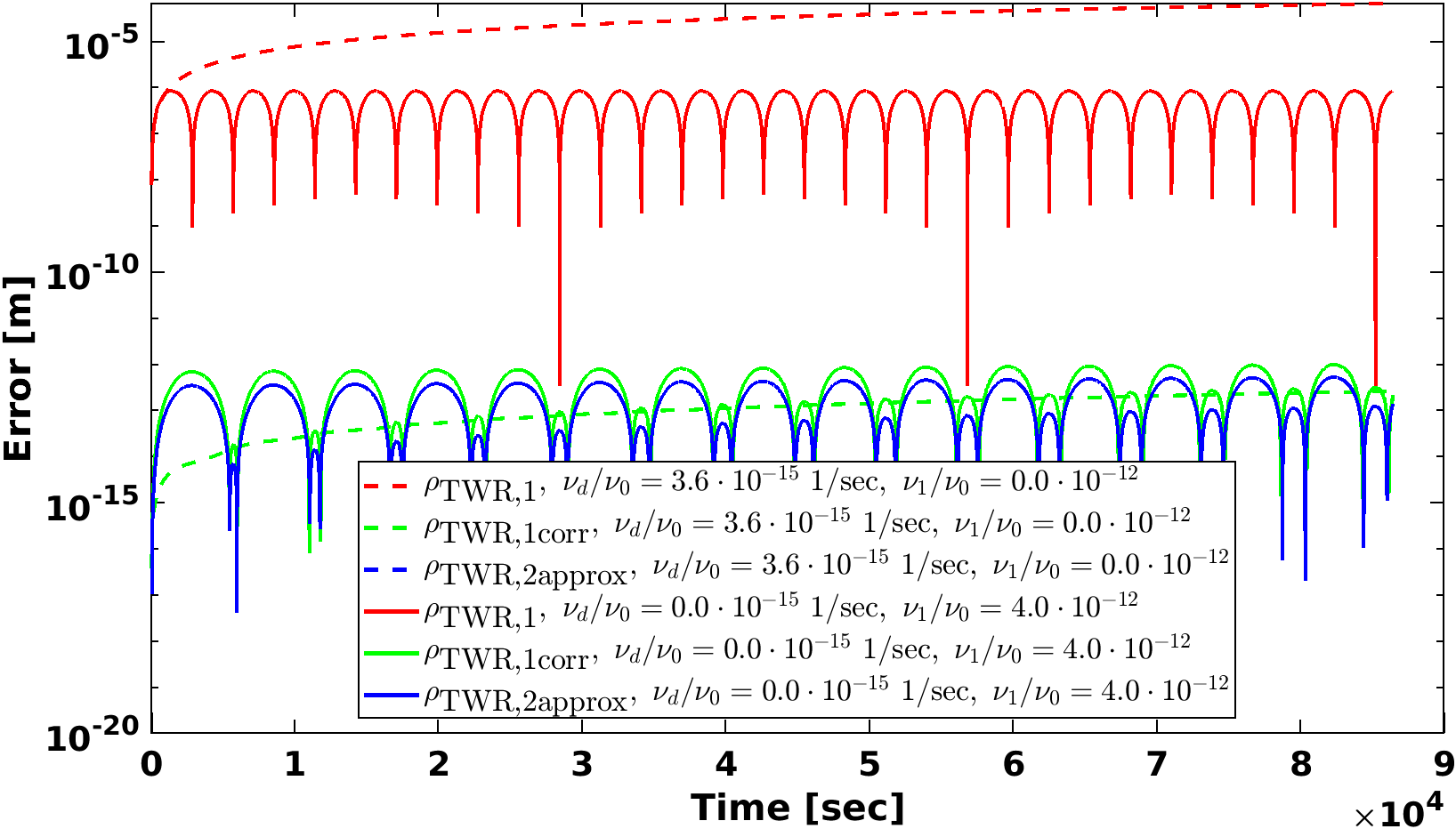}
    \caption{Evaluation of accuracy for different formulas and approximations $\rho_\textrm{TWR,...}(t)$ shown in red for eq.~(\ref{eq:Rho1}), in green for eq.~(\ref{eq:Rho1Corr1}) and in blue for eq.~(\ref{eq:dPhiApprox}). The simulation parameters for a time-series of \SI{86400}{sec} length are given in Appendix~\ref{sec:Sim} and Table~\ref{tab:SIMparam} therein. The error is defined as $|\rho_\textrm{TWR,...}(t) - (L(t)-L_0) |$, where $(L(t)-L_0)$ denotes the true (debiased) range (cf.~Appendix~\ref{sec:Sim}). The dashed traces assume a drifting frequency $\nu_\xR$, while the solid lines assume an oscillating frequency $\nu_\xR$. The dashed-blue trace is not visible because the error is zero. $\nu_\xR$ is assumed to be known without any error.}
    \label{fig:SimError}
\end{figure}

In order to assess the accuracy of the formulas given in the two previous subsections, we have used an analytical model for the optical frequency $\nu_\xR$ and the true inter-satellite distance, computed the optical phase $\Phi_\xR$ and, retrieved from that, a simulated range $\rho_\textrm{TWR}$.  The models and numerical values are given in Appendix~\ref{sec:Sim} and emulate conditions present in GRACE-FO. Based on figure~\ref{fig:SimError}, one can state that the simplistic formula without frequency correction term (eq.~(\ref{eq:Rho1})) yields a linearly drifting error of approx.~68\,\textmu m per day (dashed red trace) if the laser frequency is linearly drifting by 87\,kHz per day ($\nu_d/\nu_0 = 3.6 \cdot 10^{-15}$\,1/sec with $\nu_0=$\,282\,THz). An oscillating frequency with a fractional peak amplitude of $\nu_1/\nu_0 = 4 \cdot 10^{-12}$, which is approximately the modulation amplitude in the laser (and microwave) frequency from the proper time, yields errors in the order of 1\,\textmu m at the 1/rev oscillation frequency (solid trace).

The $\rho_\textrm{TWR,2approx}$ expression has no error when the frequency $\nu_\xR$ is drifting (dashed blue trace not visible, since zero not present in log-scale plot), and slightly lower errors than $\rho_\textrm{TWR,1corr}$ when $\nu_\xR$ is oscillating (cf.~solid blue vs. solid green trace) . However, the accuracy of the $\rho_\textrm{TWR,1corr}$ is already sufficient for GRACE-FO like applications, since the error is at the picometer level for a one day long time-series in the drifting and oscillating case. 

The error for $\rho_\textrm{TWR,2exact}$ from eq.~(\ref{eq:Rho2Corr}) cannot be shown in the log-scale figure because the equation is exact with zero error. The exact solution is our recommended way to transform the phase to range. 

\subsection{Model Errors: Scale Factor and Light Propagation Time}
\label{sec:ScaleError}
The formulas used for converting phase to range considered so far error-free knowledge of the laser frequency ($\nu_\xR$) and light travel time ($\dtRTR^\textrm{GPS}$ = \dtRTR). In order to assess the effect of inaccuracies in these models, one can replace
\begin{align}
    &\nu_\xR(t) \longrightarrow \nu_\xR(t) + \delta \nu_\xR(t) \label{eq:ReplacementRule1}\\
    &\dtRTR^\textrm{GPS}(t) \longrightarrow \dtRTR^\textrm{GPS}(t) + \dtRTR^\textrm{GPS,e}(t) \label{eq:ReplacementRule2}
\end{align}
such that $\delta \nu_\xR$ and $\dtRTR^\textrm{GPS,e}$ account for errors in the absolute laser frequency and propagation time, respectively. 
There are two ways to apply the laser frequency errors. One way is to consider them in the phase domain (cf.~eq.~(\ref{eq:varphiTWR}))
\begin{align}
    \varphi_\textrm{TWR}^\textrm{Noisy}(t) - \varphi_\textrm{TWR}^\textrm{NoiseFree}(t) &= \Phi_\textrm{R,Noise}(t) - \Phi_\textrm{R,Noise}(t - \dtRTR) = \int_{t - \dtRTR}^{t} \delta \tilde{\nu}_\xR(t^\prime)~\dd t^\prime \label{eq:LFNdirectForm} \\
    &\approx \delta \tilde{\nu}_\xR(t) \cdot \dtRTR(t),
\end{align}
where we just renamed $\delta {\nu}$ into $\delta \tilde{\nu}_\xR$. Eq.~(\ref{eq:LFNdirectForm}) illustrates that the (cavity) phase variations $\Phi_\textrm{R,Noise}$ are suppressed in the TWR combination due to the self-comparison with a delayed instance, which can be expressed in the frequency domain with a multiplier $(1-e^{-2 \pi i f \dtRTR})$. Similar equations apply to the KBR, where this multiplier is called a \emph{ranging filter} \cite[Eq.~(4.20)]{kim2000simulation} \cite[Eq.~(3.38)]{thomas1999analysis}.
The second way of applying the errors in line (\ref{eq:ReplacementRule1}) and (\ref{eq:ReplacementRule2}) is to consider them in the quantities used to convert the phase to range. When using both ways at the same time, and comparing the range with model errors ($\rho^\textrm{mod}_\textrm{...}$) to the original formula, we obtain after expansion to leading order\footnote{For the sake of simplicity, we additionally neglected terms that contain $\dtRTR^\textrm{GPS,e}$ in eq.~(\ref{eq:AdditionSimp}).}
\begin{align}
    &\rho_\textrm{TWR,1corr}^\textrm{mod} - \rho_\textrm{TWR,1corr} \nonumber \\
    \quad &\approx +\frac{c_0}{2} \tilde{\epsilon}_\textrm{SCF}(t) \cdot \dtRTR(t) -\frac{c_0}{2} \epsilon_\textrm{SCF}(t) \cdot \left(  \dtRTR(t) + \dtRTR(0) \cdot \left( \frac{ \nu_\xR(0)}{\nu_\xR(t)} - 1\right) \right) \nonumber \\ &\qquad + \frac{c_0}{2} \cdot \dtRTR^\textrm{GPS,e}(0) \cdot \left( \frac{\nu_\xR(0)}{\nu_\xR(t)} - 1\right) + \frac{c_0}{2} \cdot \epsilon_\textrm{SCF}(0) \cdot \dtRTR(0) \cdot \frac{ \nu_\xR(0)}{\nu_\xR(t)} + \textrm{const.} \\
    &\approx +\frac{c_0}{2} \tilde{\epsilon}_\textrm{SCF}(t) \cdot \dtRTR(t) - \frac{c_0}{2} \epsilon_\textrm{SCF}(t) \cdot \dtRTR(t)  + \textrm{const.} \\
   & \rho_\textrm{TWR,2approx}^\textrm{mod} - \rho_\textrm{TWR,2approx} \nonumber \\ 
   \quad &\approx -\frac{c_0}{2} \dtRTR \cdot (\epsilon_\textrm{SCF} - \tilde{\epsilon}_\textrm{SCF}) + \int_0^t \frac{c_0 \dtRTR}{2}  \cdot \left( \dtRTRdot~\frac{\delta \dot{\nu}_\xR}{\nu_\xR}  -  \epsilon_\textrm{SCF} \frac{\dot{\nu}_\xR}{\nu_\xR}\right)  \dd t^\prime + \textrm{const.} \label{eq:AdditionSimp}\\
    &\approx +\frac{c_0}{2} \tilde{\epsilon}_\textrm{SCF}(t) \cdot \dtRTR(t) - \frac{c_0}{2} \epsilon_\textrm{SCF}(t) \cdot \dtRTR(t)  + \textrm{const.},
\end{align}
where the scale factor (SCF) error was introduced in two versions as
\begin{align}
    \epsilon_\textrm{SCF}(t) = \frac{\delta \nu_\xR(t)}{\nu_\xR(t)}, \qquad \tilde{\epsilon}_\textrm{SCF}(t) = \frac{\delta \tilde{\nu}_\xR(t)}{\nu_\xR(t)}.
\end{align}
Both formulas ($\rho_\textrm{TWR,1corr}$ and $\rho_\textrm{TWR,2approx}$) and both ways to apply the error yield to leading order the same result: the scale factor error ($\epsilon_\textrm{SCF}$ and $\tilde{\epsilon}_\textrm{SCF}$) arising from the limited knowledge of the frequency $\nu_\xR$ produces ranging errors proportional to the satellite separation $c_0 \dtRTR(t) / 2 \approx L(t) $. This is the expected result and the usual coupling of frequency noise into the range measurement \cite[eq.~(11)]{sheard2012intersatellite} at least to leading order. The exact error coupling can become rather complicated.

In order to assess the magnitude of scale factor errors, we once more assume an oscillatory and a drifting behavior as in the previous subsection. For the drift, we reuse the value of $3.6 \cdot 10^{-15}$\,1/sec (cf.~table~\ref{tab:SIMparam}), which results in a scale factor change $ \Delta \epsilon_\textrm{SCF} = 3.1 \cdot 10^{-10} $ per day. Such a scale factor error is present if the cavity resonance frequency in the LRI drifts by 87\,kHz per day without knowledge about the drift, which we regard to be potentially possible. The resulting ranging error per day is then
\begin{align}
    \langle L \rangle \cdot  \Delta \epsilon_\textrm{SCF} = 220\,\textrm{km} \cdot 3.1 \cdot 10^{-10} = 68\,\textrm{\textmu m}.
\end{align}
A realistic error for the oscillatory behavior arises when the proper time induced modulation is neglected, as currently in the regular LRI data processing. This means the sinusoidal modulation with peak amplitude $\nu_\xR(t)/\langle \nu_\xR \rangle \approx 4 \cdot 10^{-12}$ at 1/rev frequency is neglected, which implies an oscillating scale factor error $\epsilon_\textrm{SCF} \approx 4 \cdot 10^{-12}$ at the orbital frequency. The resulting ranging error is
\begin{align}
    \langle L \rangle \cdot \epsilon_\textrm{SCF} \approx \SI{220}{km} \cdot 4 \cdot 10^{-12}   \approx 0.9\,\textrm{\textmu m}_\textrm{peak} \textrm{~at 1/rev frequency}. \label{eq:LriProperTimeError}
\end{align}
The calculated changes of $68\,\textrm{\textmu m}$ per day and $0.9\,\textrm{\textmu m}_\textrm{peak}$ at 1/rev frequency are actually apparent from the red traces in fig.~\ref{fig:SimError} for the error from eq.~(\ref{eq:RhoNaive}). This equation exhibits a time-dependent frequency $\nu_\xR(t)$ only in the divisor, which has very little effect on the range, as will be shown later in fig.~\ref{fig:KbrComparison}. Thus, eq.~(\ref{eq:RhoNaive}) is approximately equivalent to a formula with static or mean frequency and the dominant error in eq.~(\ref{eq:RhoNaive}) is the product $L(t=0) \cdot \epsilon_\textrm{SCF}(t) $, which is very similar to the scale factor error $\langle L \rangle \cdot \epsilon_\textrm{SCF}(t) $ discussed here.

Any static error in the scale factor, e.g. $\langle \epsilon_\textrm{SCF} \rangle = 10^{-8}$ corresponding to 2.81\,MHz frequency error of the LRI cavity, produces an error proportional to the separation $L(t)$. If one assumes a drift of 0.01\,m/s or 864\,m per day and a 1/rev amplitude of 400\,m (peak), the scale factor errors correspond to ranging errors of 8.6\,\textmu m per day and 4\,\textmu m at 1/rev frequency.

Although the derivation of the scale factor error was based on the LRI, we anticipate that the KBR instrument features a similar error coupling. The sinusoidal modulation amplitude of approx.~$4 \cdot 10^{-12}$ from the proper time is a fractional quantity and appears with the same amplitude in the microwave (cf.~sec.~\ref{sec:PhaseObservable}). Thus, the $0.9\,\textrm{\textmu m}_\textrm{peak}$ error would be applicable to the KBR as well. However, day-to-day changes in the USO frequency are usually a factor 10 or more lower than $3.1 \cdot 10^{-10}$  (cf.~left plot of fig.~\ref{fig:propertime}). Thus, the ranging error from linear drift should be at the level of a few micron level instead of $68\,\textrm{\textmu m}$ per day. In addition, static scale factor errors are expected to be much smaller in KBR, as the microwave frequency can be determined with high accuracy from GPS data in post-processing during orbit determination.

\section{KBR Dual-One Way Ranging}
\label{sec:MWI_DOWR}
In the KBR, the microwave radiation with carrier frequencies $\nu_{A/B}^{\textrm{K/Ka}}$ on satellite~A and B is produced by upconversion of the USO base frequencies $f_\textrm{A/B,USO}$ using fixed integer multipliers (see table~\ref{tab:MWIcoeff}). Two phase measurements $\varphi^\textrm{K/Ka}$ are performed per satellite, one at the K- and another at the Ka-band. These are combined into 
\begin{align}
    \varphi_\textrm{DOWR}^\textrm{K/Ka}(t) &:= \varphi_\xB^\textrm{K/Ka}(t) - \varphi_\xA^\textrm{K/Ka}(t) \\ &= \Phi_\xB^\textrm{K/Ka}(t) - \Phi_\xB^\textrm{K/Ka}(t-\Delta t_{\textrm{BA}}^\textrm{K/Ka}) + \Phi_\xA^\textrm{K/Ka}(t) - \Phi_\xA^\textrm{K/Ka}(t-\Delta t_{\textrm{AB}}^\textrm{K/Ka}) \\
    & \approx \nu_\xB^\textrm{K/Ka}(t) \cdot \Delta t_{\textrm{BA}}^\textrm{K/Ka}(t) + q_\xB^\textrm{K/Ka}  + \nu_\xA^\textrm{K/Ka}(t) \cdot \Delta t_{\textrm{AB}}^\textrm{K/Ka}(t) + q_\xA^\textrm{K/Ka}, \label{eq:DOWR1a}
\end{align}
where $\Delta t_{\textrm{AB}}^\textrm{K/Ka}(t)$ describes the light propagation time from satellite $\textrm{A}$ to $\textrm{B}$ at the respective K- or Ka frequency band at a reception time $t$. The propagation time can be expressed as
\begin{align}
    \Delta t_{\textrm{AB}}^\textrm{K/Ka}(t) = \Delta t_\textrm{media,AB}^\textrm{K/Ka}(t) +  \Delta t_{\textrm{AB}}(t) = \Delta t_\textrm{media,AB}^\textrm{K/Ka}(t) +  \Delta t_{\textrm{LTC,AB}}(t)  +  \Delta t_\textrm{inst}(t), \label{eq:tAB}
\end{align}
where the first term mainly describes the effect of the ionosphere on the propagation time, which is proportional to the squared frequency ($\Delta t_\textrm{media,AB}^\textrm{K/Ka} \propto 1/\nu_\xA^2$). The second term on the right-hand side is the light-time correction \cite{Yan2020}
\begin{align}
     \Delta t_{\textrm{LTC,AB}}(t) := \Delta t_{\textrm{AB}}(t) - \Delta t_\textrm{inst}(t)
\end{align}
and the third term is the desired instantaneous range rescaled to light propagation time
\begin{align}
    \Delta t_\textrm{inst}(t):=|\vec r_\xA(t) - \vec r_\xB(t) |/c_0. \label{eq:tinst}
\end{align}
One can easily show with the previous definitions that the linear combination \cite{kim2000simulation}
\begin{align}
    \rho_\textrm{DOWR}(t) := c_0 \cdot a^\textrm{K} \cdot \frac{ \varphi_\textrm{DOWR}^\textrm{K}(t) }{ \nu_\xA^{\textrm{K}} + \nu_\xB^{K}} + c_0 \cdot a^{\textrm{Ka}} \cdot \frac{ \varphi_\textrm{DOWR}^\textrm{Ka}(t) }{ \nu_\xA^\textrm{Ka} + \nu_\xB^\textrm{Ka}} \label{eq:DOWR1}
\end{align}
with $a^\textrm{K} = -9/7$ and $a^\textrm{Ka} = 16/7$ is a so-called ionosphere-free combination where the $\Delta t_\textrm{media}$ terms vanish. This equation is usually employed with constant frequencies $\nu_\textrm{A/B}^\textrm{K/Ka}$  in the denominator (cf.~eq.~(\ref{sec:PhaseFreqRatio})) and on daily segments in the regular KBR processing.
It is reasonable to assume that the four phase measurements $\varphi_ \textrm{A/B}^\textrm{K/Ka}(t)$ are debiased such that the phase is zero at the initial epoch $t=0$ of the time-series in order to remove the arbitrary constant. This implies that the $q$-constants in eq.~(\ref{eq:DOWR1a}) are
\begin{align}
 q_\textrm{A/B}^\textrm{K/Ka} = -\nu_\textrm{A/B}^\textrm{K/Ka}(t=0) \cdot \Delta t_\textrm{AB/BA}^\textrm{K/Ka}(t=0).   
\end{align}
Finally, by employing $q$-constants and time-dependent frequencies in eq.~(\ref{eq:DOWR1}), inserting eq.~(\ref{eq:tAB}) into eq.~(\ref{eq:DOWR1}), and considering the relation of carrier frequencies to USO frequencies from table~\ref{tab:MWIcoeff}, one obtains
\begin{align}
    \frac{ a^\textrm{K} \cdot \varphi^\textrm{K}_\textrm{DOWR}(t)}{\nu^\textrm{K}_\xA(t)+\nu^\textrm{K}_\xB(t)} + \frac{ a^\textrm{Ka} \cdot \varphi^\textrm{Ka}_\textrm{DOWR}(t)}{\nu^\textrm{Ka}_\xA(t)+\nu^\textrm{Ka}_\xB(t)} &= \Delta t_\textrm{inst}(t) - \Delta t_\textrm{DOWR,FV}(t) - \Delta t_\textrm{LTC,DOWR}(t) + \textrm{const}, \label{eq:DOWR2pre}
\end{align}
where the first term $\Delta t_\textrm{inst}$ on the right-hand side is the desired ranging signal expressed as propagation time, the second term describes the frequency variations that couple with the satellite separation ($\Delta t_\textrm{AB} \approx \Delta t_\textrm{BA} \approx L/c_0$) as
\begin{align}
    \Delta t_\textrm{FV,DOWR}(t) := \frac{\Delta t_\textrm{AB}(0) \cdot f_\textrm{A,USO}(0) + \Delta t_\textrm{BA}(0) \cdot f_\textrm{B,USO}(0) }{f_\textrm{A,USO}(t) +f_\textrm{B,USO}(t)} - \frac{\Delta t_\textrm{AB}(0) + \Delta t_\textrm{BA}(0)}{2} \label{eq:DOWR2formulaFV}
\end{align}
and the third term is the light-time correction \cite{Yan2020}
\begin{align}
    \Delta t_\textrm{LTC,DOWR}(t) := \Delta t_\textrm{LTC,AB}(t) \cdot b_\textrm{AB}(t) + \Delta t_\textrm{LTC,BA}(t) \cdot b_\textrm{BA}(t),
\end{align}
where the coefficients $b_\textrm{AB}$ and $b_\textrm{BA}$ are given in table~\ref{tab:MWIcoeff}. The second and third term on the right-hand side of eq.~(\ref{eq:DOWR2pre}) show the needed correction terms in order to obtain the range independent of USO frequency variations, i.e.
\begin{align}
    \rho_\textrm{KBR} &:= c_0 \cdot \left( \frac{ a^\textrm{K} \cdot \varphi^\textrm{K}_\textrm{DOWR}}{\nu^\textrm{K}_\xA+\nu^\textrm{K}_\xB} + \frac{ a^\textrm{Ka} \cdot \varphi^\textrm{Ka}_\textrm{DOWR}}{\nu^\textrm{Ka}_\xA+\nu^\textrm{Ka}_\xB} + \Delta t_\textrm{FV,DOWR} + \Delta t_\textrm{LTC,DOWR}  + \Delta t_\textrm{AOC} \right), \label{eq:DOWR2}
\end{align}
where only $c_0$ and $a_\textrm{K/Ka}$ are time-independent.

The instantaneous biased range $\rho_\textrm{KBR}$ can be directly employed in gravity field recovery. The so-called antenna offset correction $\rho_\textrm{AOC}$ was added to be consistent with the official SDS data processing and is discussed in sec~\ref{sec:TTL}.

The light-propagation times $\Delta t_\textrm{AB}$ and $\Delta t_\textrm{BA}$ needed to compute the frequency correction term in eq.~(\ref{eq:DOWR2formulaFV}) can be obtained from the GPS orbit products and are also used for the light-time correction computation. The quantities differ by $c_0|\Delta t_\textrm{AB} - \Delta t_\textrm{BA}|  \approx \SI{10}{m}$ in GRACE-FO due to the finite speed of light and the relative velocity between satellites \cite{Yan2020}. Note that the USO frequencies $ f_\textrm{A/B,USO}$ in eq.~(\ref{eq:DOWR2formulaFV}) can be replaced with the carrier frequency of the K- or the Ka-band, since
\begin{align}
    \frac{f_\textrm{A/B,USO}(0)}{f_\textrm{A/B,USO}(t)} = \frac{\nu_\textrm{A/B}^\textrm{K}(0)}{\nu_\textrm{A/B}^\textrm{K}(t)} = \frac{\nu_\textrm{A/B}^\textrm{Ka}(0)}{\nu_\textrm{A/B}^\textrm{Ka}(t)}.
\end{align}

Compared to previous derivations and the official SDS processing described in \cite{GRACE_Level1_Handbook}, the novel aspect of eq.~(\ref{eq:DOWR2})  is that it considers a time-variable frequency when converting the phase to range, and it contains a correction term $\Delta t_\textrm{FV,DOWR}$ for the frequency variations. In the following sections, we show that this term might be of relevance when LRI and KBR data is compared at low frequencies.



\begin{table}[H] 
	\caption{Numerical values for frequencies and coefficients used to describe dual one-way ranging of the KBR. $\dagger$: assumes $\dd \tau^\textrm{USO}_\xA / \dd t = \dd \tau^\textrm{USO}_\xB / \dd t  = 1$. }
	\label{tab:MWIcoeff}
\begin{tabularx}{\textwidth}{p{1.2cm}p{6.5cm}p{3.5cm}}
\toprule
\textbf{Name}	& \textbf{Value/Formula}	& \textbf{Comment}\\
\midrule
		$\hat{f}_\textrm{A,USO}$   &  \SI{4.832000}{MHz} & exact  \\
		$\hat{f}_\textrm{B,USO}$   &  \SI{4.832099}{MHz} & exact  \\
		${f}_\textrm{A,USO}(t)$   &  $\SI{4.832000}{MHz} \cdot \dd \tau^\textrm{USO}_\xA / \dd t$ &   \\
		${f}_\textrm{B,USO}(t)$   &  $\SI{4.832099}{MHz} \cdot \dd \tau^\textrm{USO}_\xB / \dd t$ &   \\		
		$\nu_\xA^\textrm{K}(t)$   & $f_\textrm{A,USO}(t) \cdot 5076$  &  $\SI{24.527232}{GHz}^\dagger$  \\
		$\nu_\xA^\textrm{Ka}(t)$   & $f_\textrm{A,USO}(t)\cdot 6768$  &  $\SI{32.702976}{GHz}^\dagger$   \\
		$\nu_\xB^\textrm{K}(t)$   & $f_\textrm{B,USO}(t) \cdot 5076$  &  $\SI{24.527734524}{GHz}^\dagger$   \\
		$\nu_\xB^\textrm{Ka}(t)$   & $f_\textrm{B,USO}(t) \cdot 6768$  & $\SI{32.703646032}{GHz}^\dagger$  \\
		$a^\textrm{K}$   & $\frac{-\nu_\xA^\textrm{K} \cdot \nu_\xB^\textrm{K}}{(\nu_\xA^\textrm{Ka} \cdot \nu_\xB^\textrm{Ka}-\nu_\xA^\textrm{K} \cdot \nu_\xB^\textrm{K})} = -9/7$ & exact   \\
		$a^\textrm{Ka}$  & $\frac{\nu_\xA^\textrm{Ka} \cdot \nu_\xB^\textrm{Ka}}{(\nu_\xA^\textrm{Ka} \cdot \nu_\xB^\textrm{Ka}-\nu_\xA^\textrm{K} \cdot \nu_\xB^\textrm{K})} = 16/7$ &  exact  \\
		$b_\textrm{AB}^\textrm{K}(t)$  & $\frac{(\nu_\xA^\textrm{K})^2 \cdot \nu_\xB^\textrm{K}}{(\nu_\xA^\textrm{K}+\nu_\xB^\textrm{K}) (\nu_\xA^\textrm{K} \nu_\xB^\textrm{K} - \nu_\xA^\textrm{Ka} \nu_\xB^\textrm{Ka})} = \frac{-9 {f}_\textrm{A,USO}(t)}{7 \cdot ({f}_\textrm{A,USO}(t) +{f}_\textrm{B,USO}(t) ) }$ & $\frac{-43488000}{67648693}\approx -0.642851^\dagger $  \\
		$b_\textrm{AB}^\textrm{Ka}(t)$ & $- \frac{(\nu_\xA^\textrm{Ka})^2 \cdot \nu_\xB^\textrm{Ka}}{(\nu_\xA^\textrm{Ka}+\nu_\xB^\textrm{Ka}) (\nu_\xA^\textrm{K} \nu_\xB^\textrm{K} - \nu_\xA^\textrm{Ka} \nu_\xB^\textrm{Ka})} = \frac{16 {f}_\textrm{A,USO}(t)}{7 \cdot ({f}_\textrm{A,USO}(t) +{f}_\textrm{B,USO}(t) ) }$ &  $\frac{77312000}{67648693}\approx 1.1428454^\dagger $ \\
		$b_\textrm{BA}^\textrm{K}(t)$  & $\frac{\nu_\xA^\textrm{K} \cdot (\nu_\xB^\textrm{K})^2}{(\nu_\xA^\textrm{K}+\nu_\xB^\textrm{K}) (\nu_\xA^\textrm{K} \nu_\xB^\textrm{K} - \nu_\xA^\textrm{Ka} \nu_\xB^\textrm{Ka})} = \frac{-9 {f}_\textrm{B,USO}(t)}{7 \cdot ({f}_\textrm{A,USO}(t) +{f}_\textrm{B,USO}(t) ) }$ &  $\frac{-43488891}{67648693}\approx -0.642864^\dagger $ \\
		$b_\textrm{BA}^\textrm{Ka}(t)$ & $- \frac{\nu_\xA^\textrm{Ka} \cdot (\nu_\xB^\textrm{Ka})^2}{(\nu_\xA^\textrm{Ka}+\nu_\xB^\textrm{Ka}) (\nu_\xA^\textrm{K} \nu_\xB^\textrm{K} - \nu_\xA^\textrm{Ka} \nu_\xB^\textrm{Ka})} = \frac{16 {f}_\textrm{B,USO}(t)}{7 \cdot ({f}_\textrm{A,USO}(t) +{f}_\textrm{B,USO}(t) ) }$ & $\frac{77313584}{67648693}\approx 1.142869^\dagger $  \\
		$b_\textrm{AB}(t)$  & $b_\textrm{AeBr}^\textrm{K}+b_\textrm{AeBr}^\textrm{Ka} = \frac{ {f}_\textrm{A,USO}(t)}{{f}_\textrm{A,USO}(t) +{f}_\textrm{B,USO}(t)}$ &  $\approx 0.499995^\dagger$ \\ 
		$b_\textrm{BA}(t)$  & $b_\textrm{BeAr}^\textrm{K}+b_\textrm{BeAr}^\textrm{Ka}= \frac{ {f}_\textrm{B,USO}(t)}{{f}_\textrm{A,USO}(t) +{f}_\textrm{B,USO}(t)}$ &  $\approx 0.500005^\dagger$ \\  
\bottomrule
\end{tabularx}
\end{table}
\unskip

\section{Range Differences between LRI and KBR}
\label{sec:KBRLRIresid}
The LRI and KBR ranging data can be directly compared at the instantaneous range level, meaning that light-time corrections of KBR and LRI and KBR antenna offset correction are applied. The light-time correction is different for KBR DOWR and LRI TWR. The ionospheric effect is already removed by default in KBR1B datasets, but the correction is provided separately in case one wants to derive the variations in electron content of the ionosphere. Since KRB1B data is provided at a rate of \SI{0.2}{Hz} and LRI1B data at a rate of \SI{0.5}{Hz}, the data has to be resampled, ideally through low-pass filtering and decimation in order to avoid aliasing.

The LRI data as provided in the official v04 LRI1B dataset has already been deglitched, which is a process that removes steps and glitches in the ranging data that are correlated with some thruster activations \cite{abich2019orbit}. Moreover, the LRI ranging data is rescaled and time-shifted such that KBR-LRI residuals are reduced. This rescaling is necessary due to the LRI absolute laser frequency $\nu_\xR$, defining the conversion factor from phase to range, only being known with low accuracy in GRACE-FO. In contrast to the KBR frequencies $\nu_\textrm{A/B}^\textrm{K/Ka}$, the laser frequency $\nu_\xR$ cannot be directly measured in-flight on GRACE-FO, although attempts to derive the frequency from laser telemetry are being made \cite{misfeldt2021thermal}. However, future missions will likely host a LRI scale factor unit that enable measurements of a time-resolved frequency \cite{Rees:21}.

The deglitching and rescaling of LRI data was sub-optimal in v04 LRI1B data before 2020-06-30, but improved significantly thereafter due to changes in processing \cite{ReleaseNotesLevel1}. 

\begin{figure}
\begin{adjustwidth}{-\extralength}{0cm}
    \centering
    \includegraphics[width=8.5cm]{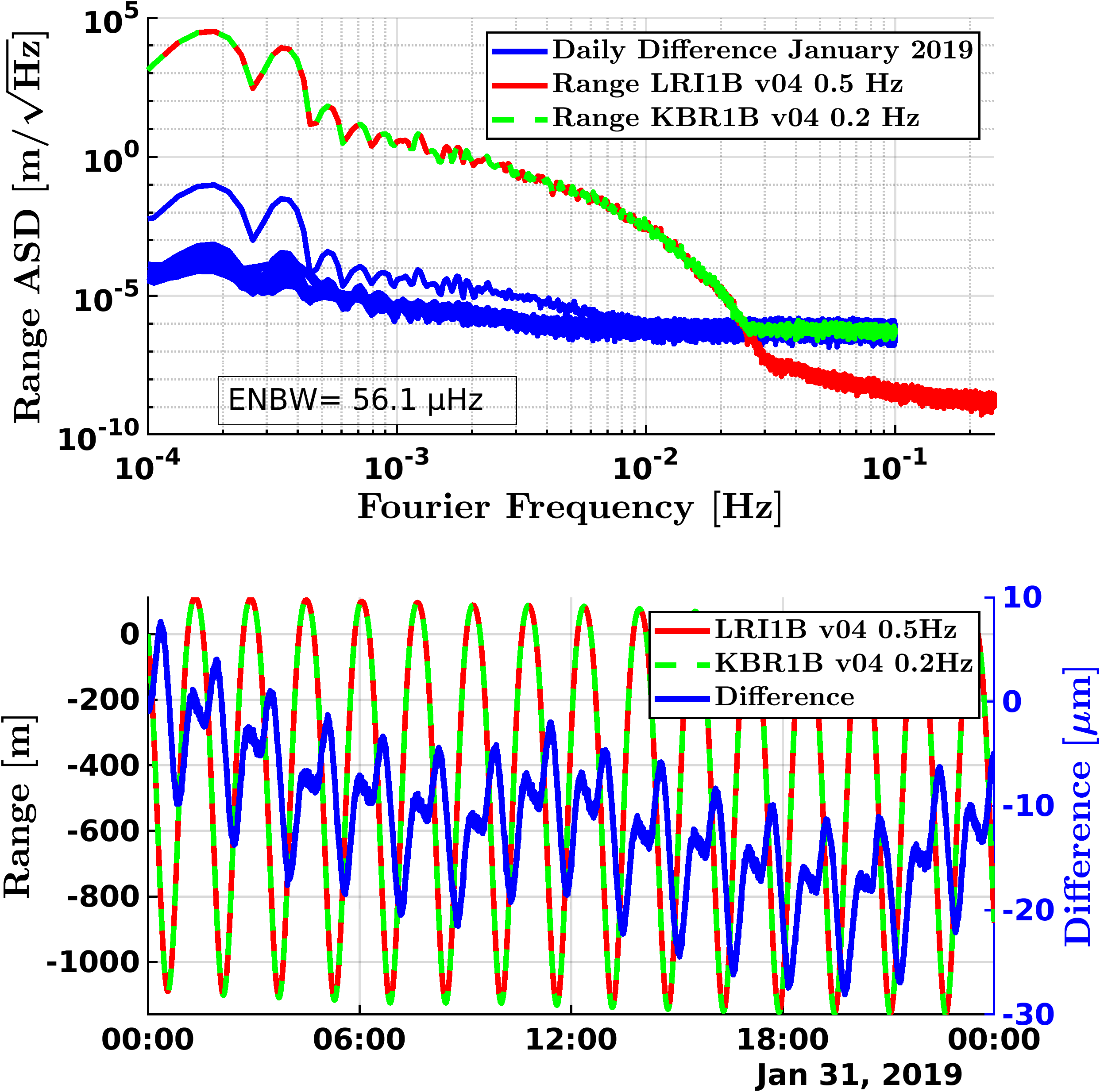}~\includegraphics[width=8.5cm]{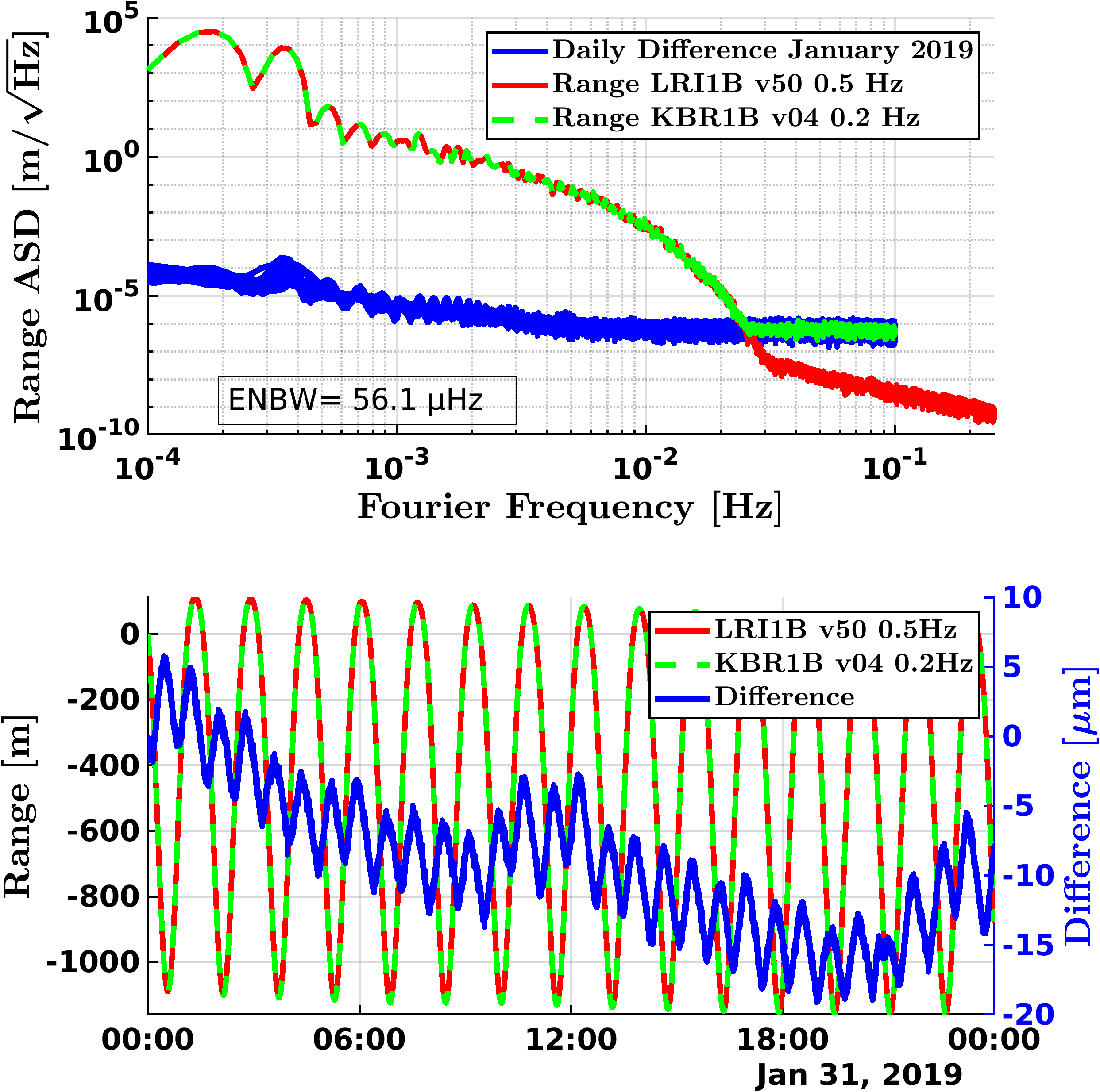}
\end{adjustwidth}    
    \caption{Typical LRI and KBR ranging data shown in the spectral domain (top) and in the time-domain (bottom). The left plots use the official LRI1Bv04 data product, while the right plots show an in-house derived alternative product (LRI1Bv50). }
    \label{fig:ResADSs}
\end{figure}

We previously derived an alternative LRI1B product from level1a data, which we label as v50 in order to distinguish it from the official v04 product \cite{lmuller2020alternative}. A comparison of both datasets is shown for January 2019 in figure~\ref{fig:ResADSs}. Other time-periods exhibit the same characteristics. The range signal in GRACE-FO has dominant 1/rev (and 2/rev) oscillations of several \SI{100}{m} amplitude and a continuous spectrum containing the interesting gravity field information and non-gravitational accelerations up to approx.~\SI{35}{mHz}. At high frequencies, the LRI and KBR range data is dominated by the respective instrument noise.

The traces showing the difference between KBR and LRI reveal that these residuals are dominated by KBR noise at frequencies above \SI{35}{mHz} and the residual level increases towards lower frequencies. The residuals show pronounced 2/rev oscillations in LRI1Bv04 and LRI1Bv50 with a few micron amplitude. However, LRI1Bv04 has an additional 1/rev oscillation. When we re-estimate a scale and time-shift of LRI1Bv04, this 1/rev peak vanishes in the residuals and the results get closer to LRI1Bv50. This is an indication of the scale and time-shift being slightly inaccurate in the official v04 data. Therefore, we will often show the original LRI1Bv04 data as well as our rescaled and time-shifted version without the 1/rev oscillations in the following.

The day 2019-01-16 is visible as an outlier in the residuals (upper left plot of fig.~\ref{fig:ResADSs}), because the deglitching did not properly remove a glitch in the LRI1Bv04 data, which then disturbed the time-shift estimation.

\section{Instrument Noises and Corrections}
\label{sec:InstrumentNoises}

\begin{figure}
\begin{adjustwidth}{-\extralength}{0cm}
    \centering
    \includegraphics[width=15.9cm]{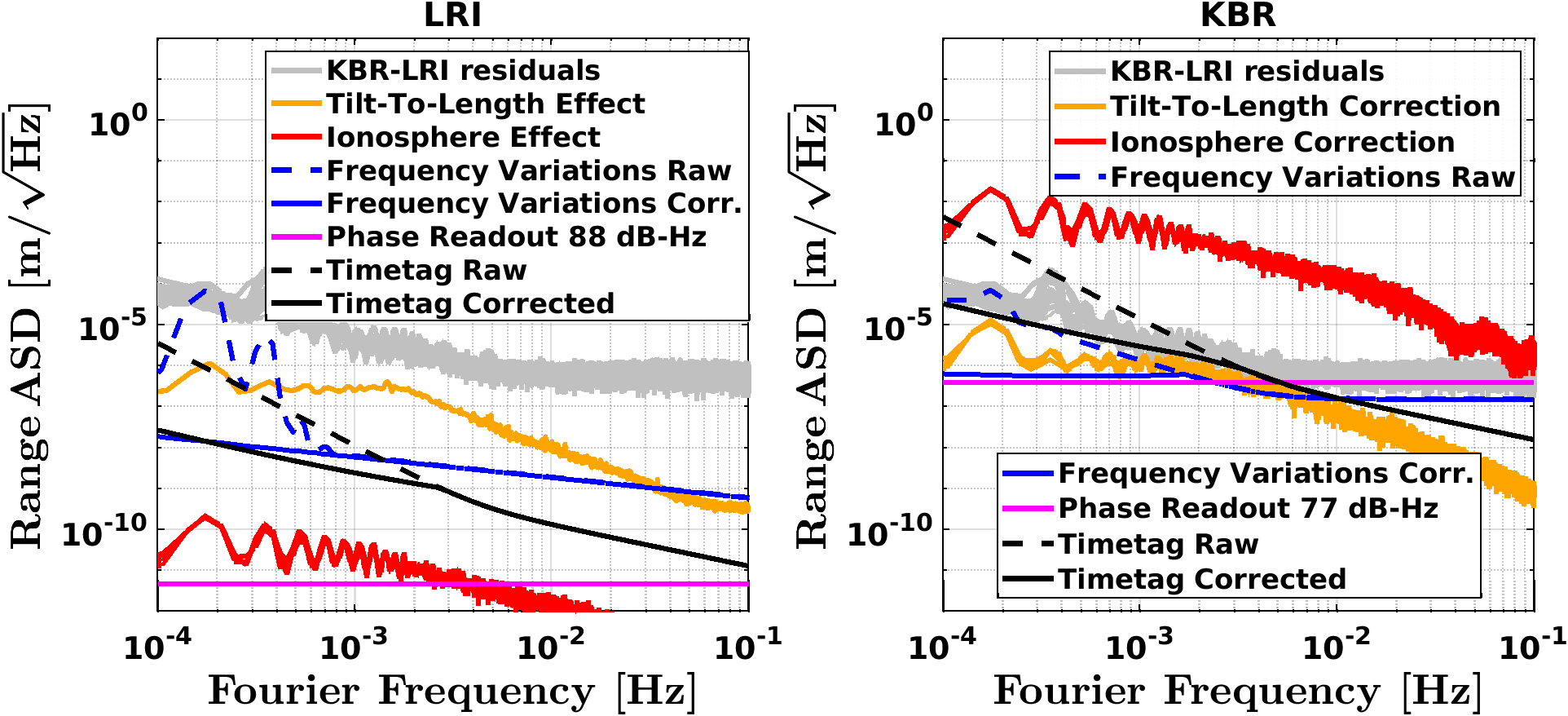}
\end{adjustwidth}
    \caption{Noise contributors and corrections in LRI and KBR in terms of amplitude spectral densities (ASDs). The grey KBR-LRI residuals are described in sec.~\ref{sec:KBRLRIresid} and replotted from fig.~\ref{fig:ResADSs}. The KBR tilt-to-length correction (Antenna-Offset-Correction) in orange is provided in the KBR1B dataset and already removed in the KBR range, while the corresponding effect in the LRI has a lower magnitude and was derived based on \cite{wegener2020} - see sec.~\ref{sec:TTL} for more details. The ionospheric correction (red) is relevant for KBR, but negligible in LRI due to the scaling with wavelength. The smaller wavelength of the laser scales the phase readout noise (magenta) to a negligible level as well, while in KBR it is dominant at high frequencies (cf.~sec.~\ref{sec:PhaseReadoutNoise}). Raw frequency variations (dashed blue) assume a static frequency (within one day) and are related to the stability of the optical cavity and proper time effect in the LRI, whereas they are related to the USO stability in KBR. These raw frequency variations can be corrected by using a time-dependent frequency (cf.~sec.~\ref{sec:CarrierFreqVar}). The effect of timetag errors is being corrected in LRI and KBR by employing CLK1B \epstime~information (dashed vs. solid black traces). For some legend entries, multiple days were plotted in order to illustrate the stationarity and variability of the traces. The mean satellite separation is $ L = \SI{194}{km}$ for January 2019. The ENBW for both plots is 52\,\textmu Hz.}
    \label{fig:NoiseComp}
\end{figure}

In order to understand which noises or error sources are limiting the KBR-LRI residuals, we have plotted many of the known effects in KBR and LRI in fig.~\ref{fig:NoiseComp}. 

We have analyzed the light-time correction extensively in \cite{Yan2020} and skip it therefore in our analysis given here, as we found no indication that the accuracy of this correction could be limiting KBR-LRI residuals in our previous work.

\subsection{Ionospheric correction}
The propagation of electromagnetic waves like microwaves or laser light is affected by the non-ideal vacuum, in particular by charged particles forming the ionosphere. The KBR system uses the dual-band observations to measure and correct the effect. The magnitude of this correction, as provided in the KBR1Bv04 data, is shown as red trace in the right plot of fig.~\ref{fig:NoiseComp}. The correction is well above the KBR-LRI residuals and the accuracy of the correction could in principle cause some error. However, we assessed the second-order ionospheric effect \cite{hernandez2007second} and concluded that the ionosphere is likely not limiting KBR-LRI residuals, since the second-order effect is direction-dependent and highly rejected in the DOWR combination, in which the contribution of the path from satellite A to B is counteracted by the contribution of the path from B to A.

The ionospheric effect as measured by the KBR can be rescaled to the LRI optical wavelength of \SI{1064.5}{nm}, see red trace on left plot of fig.~\ref{fig:NoiseComp}. For such near-infrared wavelengths, the ionospheric delay is a negligible error contributor  \cite{sheard2012intersatellite}.

\subsection{Tilt-To-Length Coupling}
\label{sec:TTL}
Satellite rotations cause pathlength changes when the reference points (RPs) of the LRI or the antenna phase centers (APCs) of KBR are not co-located with the respective center-of-mass of satellites, which are the pivot points for rotations in space \cite[sec.~2.6.3]{mueller2017design}. The RPs and APCs are effectively the fiducial points used to determine the biased inter-satellite range from interferometric phase readout and are fixed in the satellite body frame, i.e. moving in inertial space when the satellites rotate.

The main effect of the coupling can be expressed in terms of a vector describing the offset between center-of-mass and the RP or APC that is projected onto the line-of-sight as the measurement axis. To first order, lateral components perpendicular to the line-of-sight ($y$ \& $z$) couple linearly with yaw and pitch, i.e.~100\,\textmu m offset produce a yaw and pitch coupling of 100\,\textmu m/rad per satellite into the biased range. The $x$-offset couples quadratically with yaw and pitch as 100\,\textmu m/$\textrm{rad}^2$.

In the KBR context, the tilt-to-length coupling is called antenna offset correction \cite{GFO_Level1_Handbook,GRACE_Level1_Handbook} and it is provided in the KRB1B data set in terms of the DOWR-range correction, including effects from both spacecraft (cf.~$c_0 \Delta t_\textrm{AOC}$ in eq.~(\ref{eq:DOWR2})). Since the KBR antenna is at the front panel of the satellite, the APCs are offset by approx.~\SI{1.47}{m} in $x$ direction from the center-of-mass. This means that the correction contains a significant quadratic contribution from the pointing angles.
The coupling factors can be measured in-orbit using so-called KBR calibration maneuver during which the satellites are deliberately rotated.

In the LRI, the RPs on both satellites were co-located to the center-of-mass to the level of the mechanical integration accuracy in the order of 100\,\textmu m in $y$ and $z$, which yields the dominant linear contribution. The LRI tilt-to-length coupling factors can be measured using the center-of-mass calibration maneuvers \cite{wegener2020} that are usually repeated every six months. As of version 04 of LRI1B data, the tilt-to-length correction is not derived or applied to LRI data, but this may change in future releases. More information in regard to LRI tilt-to-length coupling can be found in \cite{wegener2022}.

The magnitude of the tilt-to-length effect on the KBR and LRI range measurement is shown in orange in fig.~\ref{fig:NoiseComp}. In the LRI case, the effect is approximately one order of magnitude smaller compared to the KBR. Based on these results, we can conclude that the tilt-to-length effect does not limit the KBR-LRI residuals.

\subsection{Phase Readout Noise}
\label{sec:PhaseReadoutNoise}
\begin{figure}
\begin{adjustwidth}{-\extralength}{0cm}
    \centering    
    \includegraphics[width=8.8cm]{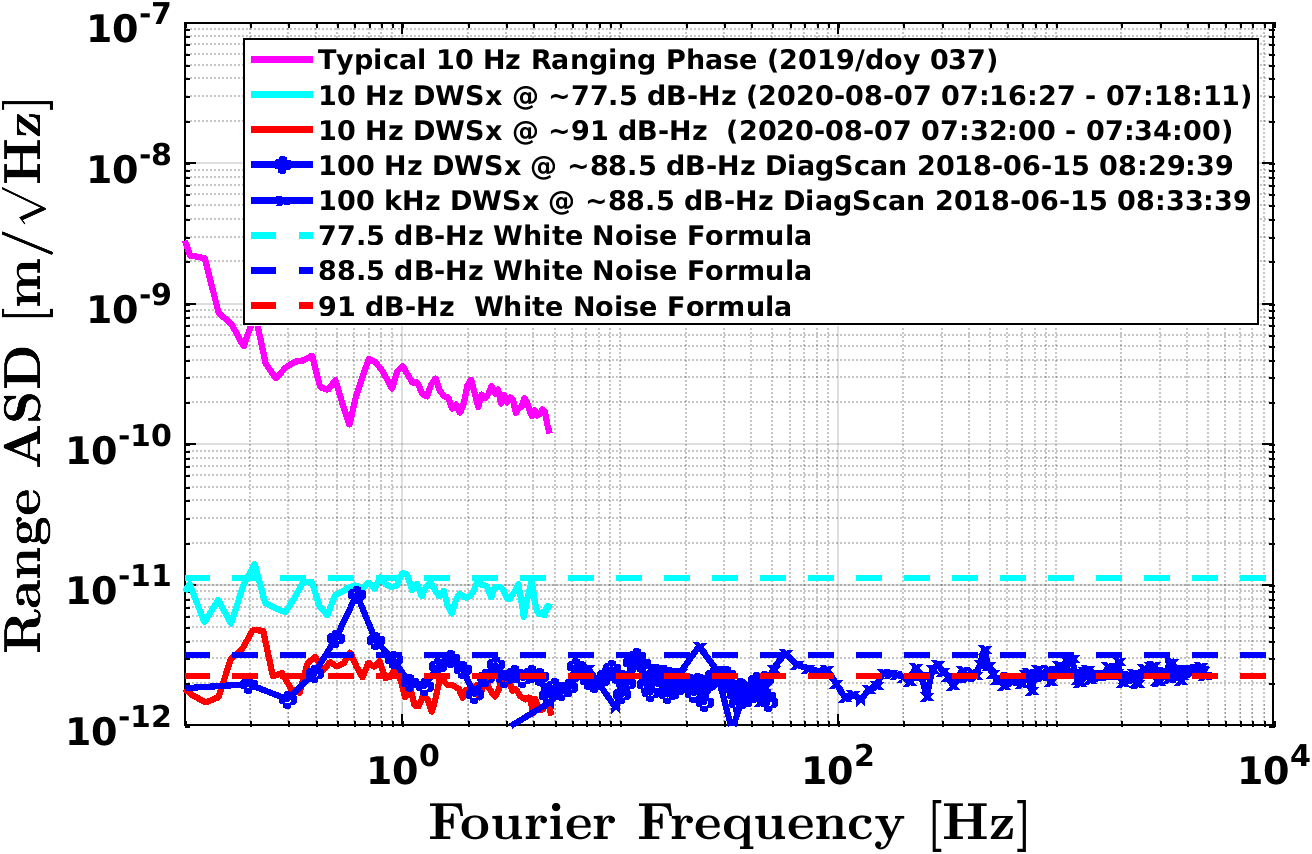}
    \includegraphics[width=8.8cm]{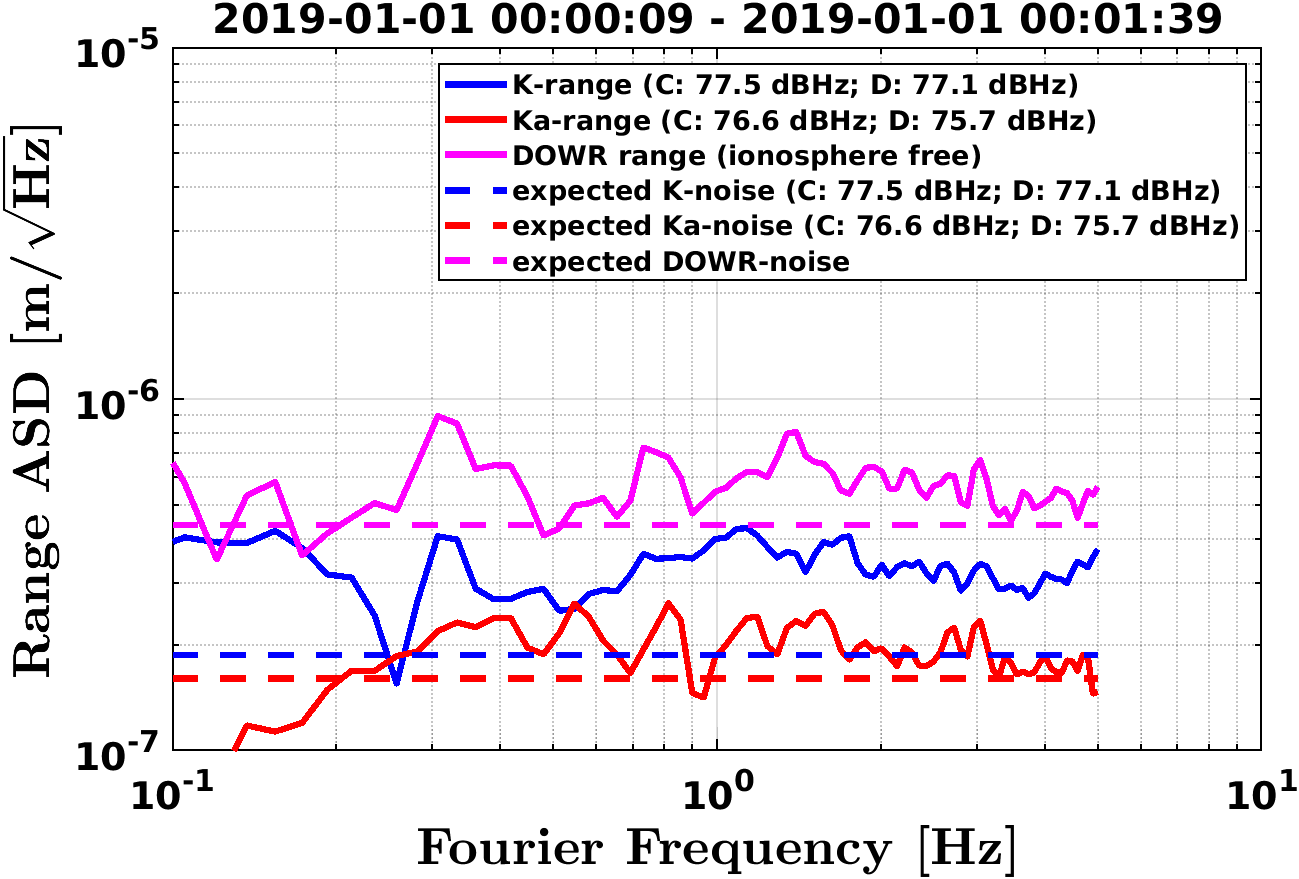}
\end{adjustwidth}            
    \caption{CNR-limited phase readout noise in LRI (left) and KBR (right). In the LRI, the phase readout noise is well below the actual ranging signal, which is limited by laser frequency noise (magenta, left plot). In KBR, the phase readout noise is likely limiting the KBR range at frequencies above \SI{10}{mHz}. The measured noise level in the K-band is slightly higher than the expected one derived from the CNR-values. This also yields a small discrepancy in the final ionosphere-free DOWR combination (magenta, dashed $0.4\,\textrm{\textmu m}/\sqrt{\textrm{Hz}}$ vs. solid $\approx 0.6\,\textrm{\textmu m}/\sqrt{\textrm{Hz}}$).}
    \label{fig:CrnNoise}
\end{figure}
The KBR and LRI both track the phase of an interference beatnote that is present as an electrical signal in the corresponding processing unit. Phase changes are proportional to changes in the light or microwave propagation time up to first order and, thus, to distance changes between spacecraft. The readout noise of the phase is usually determined by the carrier-to-noise ratio (CNR) entering the phase-tracking loop. which can be expressed in terms of a ranging error as~\cite{mueller2017design}
\begin{align}
    \textrm{ASD}[\rho_\textrm{CNR}](f) = \frac{\lambda \cdot \SI{1}{rad}}{2 \cdot 2 \pi \cdot \sqrt{\textrm{CNR}}}, \quad \textrm{CNR} = \frac{I_\textrm{rms}^2}{\textrm{PSD}[\mathcal{N}](f_\textrm{beat})}, \quad [ \textrm{ASD}[\rho_\textrm{CNR}] ] = \SI{}{m/\sqrt{Hz}}, \label{eq:CNRsimple}
\end{align}
where the half-wavelength $\lambda/2$ is used to rescale the white phase noise level to units of half-round-trip range.

The CNR is defined as the ratio of root-mean-square (rms) carrier power $I_\textrm{rms}^2$, i.e.~squared rms amplitude of the beatnote photocurrent, to the noise power spectral density ($\textrm{PSD}[\mathcal{N}]$) of the photocurrent at the beatnote frequency. Typical values for the LRI are above $\SI{80}{dBHz} = 10^8\,\SI{}{Hz}$ for both the reference and transponder satellite, which is well above the minimum requirement of $\SI{70}{dBHz} = 10^7\,\SI{}{Hz}$ of the instrument. The transponder phase measurement is used as a sensor in a feedback control loop to control the transponder laser frequency with high gain and bandwidth, e.g.~the phase variations in terms of range on the transponder side are below $\SI{10}{pm/\sqrt{\textrm{Hz}}}$ at \SI{1}{Hz} and even smaller for lower frequencies. Thus, it is an in-loop measurement close to zero. Any phase variations arising on the transponder, e.g.~from laser noise or readout noise, are imprinted onto the laser beam and transmitted to the reference satellite. Hence, the reference satellite measures the phase readout noise of transponder and reference satellite
\begin{align}
        \textrm{ASD}[\rho_\textrm{LRI,CNR}](f) = \frac{\lambda_\xR \cdot \SI{1}{rad} }{2 \cdot 2 \pi} \cdot \sqrt{\frac{1}{\textrm{CNR}_\xT}+ \frac{1}{\textrm{CNR}_\xM} }.
\end{align}
This formula is actually independent of the loop gain of the frequency-controller, i.e.~even with low gain and significant signal in the transponder phase, the ranging signal $\rho_\textrm{LRI}$ is usually formed by the combination of both satellites in order to remove the common phase ramp (cf.~eq.~(\ref{eq:varphiTWR})) and to recover the complete ranging signal. Thus, it will exhibit the phase readout noise of both satellites.

The CNR-limited phase readout noise $\textrm{ASD}[\rho_\textrm{CNR}]$ is, strictly speaking, non-stationary, as the CNR is time-dependent. Therefore, we plot short segments of data in order to prevent potential artifacts due to non-stationarity. The left plot of fig.~\ref{fig:CrnNoise} shows the calculated phase readout noise compared to the ranging signal of the LRI (magenta trace).
A CNR value of $\SI{77.5}{dBHz} = 10^{7.75}\,\SI{}{Hz}$ from a single satellite yields approx.~$\SI{11}{pm/\sqrt{Hz}}$ of white phase readout noise, or $\sqrt{2} \cdot \SI{11}{pm/\sqrt{Hz}}$ from two satellites. Such low values are not observable in the ranging phase of the LRI (mangenta trace in left plot of fig.~\ref{fig:CrnNoise}) due to other noise sources, mainly laser frequency variations from the cavity.

However, the LRI system measures the interference beatnote on a satellite with four channels in order to make use of the Differential Wavefront Sensing (DWS) technique \cite{sheard2012intersatellite}. The ranging information is given by the average of the four phase measurements, which is the linear combination of four channels with multipliers $+ 1/4 $. The CNR values reported by the LRP of the LRI refer to the coherent sum (average) of four channels. However, if the noise among different channels is uncorrelated and the phase difference among different channels is small (cf.~\cite[sec.~2.6.9]{mueller2017design} for details), the same CNR values would apply to other linear combinations as well. These other three linear combinations with multipliers $\pm 1/4$ can be called DWSyaw, DWSpitch and DWSx. DWSyaw and DWSpitch represent the angular misalignment between the interfering laser beams, and are actually zeroed by a feedback control loop using a fine-steering mirror as actuator. The remaining DWSx combination is not zeroed and exhibits white noise behavior for frequencies above \SI{0.1}{Hz} (cf.~left plot in fig.~\ref{fig:CrnNoise}) with a magnitude consistent with the expected CNR-limited phase readout noise. It is noteworthy that DWSx is a local quantity that needs to be compared against the single spacecraft CNR-limit (eq.~(\ref{eq:CNRsimple})). If noise is transferred between satellites via the laser light, it will affect all channels on the receiver, thus appearing in the average phase of the four channels, but not in the DWS combinations of the receiver.

The consistency between DWSx noise level and analytical CNR-limited phase readout noise level is at least an indicator that the CNR-limited phase readout noise is at the expected level in the ranging signal (magenta trace), though it is not directly measurable.

The phase readout noise in the context of the KBR is usually called system noise \cite{thomas1999analysis, kim2000simulation}. Based on eq.~(\ref{eq:CNRsimple}) and the DOWR combination discussed in sec.~\ref{sec:MWI_DOWR}, it is straightforward to express the CNR-limited phase readout noise in the ionosphere-free DOWR range as
\begin{align}
        \textrm{ASD}[\rho_\textrm{KBR,CNR}](f) = \frac{\SI{1}{rad}}{2 \cdot 2 \pi} \cdot \sqrt{ \frac{(a^\textrm{K} \lambda_\xA^\textrm{K})^2}{\textrm{CNR}_\xA^\textrm{K}}+ \frac{(a^\textrm{K} \lambda_\xB^\textrm{K})^2}{\textrm{CNR}_\xB^\textrm{K}}  +  \frac{(a^\textrm{Ka} \lambda_\xA^\textrm{Ka})^2}{\textrm{CNR}_\xA^\textrm{Ka}}+ \frac{(a^\textrm{Ka} \lambda_\xB^\textrm{Ka})^2}{\textrm{CNR}_\xB^\textrm{Ka}} },
\end{align}
where coefficients $a^\textrm{K}$ and $a^\textrm{Ka}$ as well as wavelength $\lambda = c_0 / \nu$ can be obtained from table~\ref{tab:MWIcoeff}. This expression is consistent with the derivation given in \cite[sec.~B-1]{thomas1999analysis} and shown as magenta-dashed trace in the right plot of fig.~\ref{fig:CrnNoise}. Unfortunately, the actual DOWR ranging noise is slightly higher than predicted based on the CNR values reported in the KBR1B data product. When plotting the individual range measurements at K- and Ka-band, we can identify that the discrepancy is caused by the K-band range, which exhibits slightly higher noise than predicted by the CNR. Though the reason of this discrepancy is unclear to the authors of this paper, the deviation is rather small. 
Based on fig.~\ref{fig:NoiseComp}, we conclude that the phase readout noise is negligible for the LRI, while it likely limits the KBR and therefore the KBR-LRI residuals at high frequencies to a white noise level of approx.~0.6\,\textmu m/$\sqrt{\textrm{Hz}}$.

\subsection{Timetag Errors}
\label{sec:TimetagError}
The telemetry of KBR and LRI contains recorded instrument timetags, which are equally sampled at their respective nominal rate. Timetag errors arise when the timetags in the instrument timeframe are converted to another time system, usually to GPS time.

The KBR and LRI are driven by the same ultra-stable oscillator (USO). Thus, the KBR time, which is often called IPU receiver time, and the LRI time are both realizations of the USO time, but KBR/IPU and LRI time have a quasi-static offset relative to each other. It is quasi-static, because it changes whenever one of the instruments reboots. And the offset is usually less than a second, since the instruments synchronize their respective time to the integer GPS second after the reboot occurred.

Since KBR and GPS measurements are performed within the IPU, the KBR timetags can be directly converted to GPS time using the \epstime~value from the CLK1B product (cf.~eq.~(\ref{eq:EpsTimeDef})) that is determined during precise orbit determination for each satellite.

For the LRI, the conversion requires addtional steps. First, datation reports as given in the LHK1A/B data product provide the timetag difference between LRI time and OBC time. Second, the TIM1B data product contains the offset between OBC time and KBR/IPU time, which is virtually zero during nominal operation, as the OBC time is steered towards the IPU time. In the end, the \epstime~from the CLK1B data product yields the final relation needed to convert the timetag to GPS time.

In fig.~\ref{fig:NoiseComp}, the curves labeled \emph{timetag raw} shown in dashed black indicate a hypothetical noise calculated under the assumption of using a daily mean value of \epstime, e.g.~no \epstime~information is employed for frequencies above $\approx 1/\SI{86400}{s}$. Thereby, the error in the timetag conversion from instrument to GPS time would be given by the variability of \epstime. The in-flight variability was plotted as spectral density in the right plot of fig.~\ref{fig:propertime} for GF-1 (blue) and GF-2 (cyan blue). The curves increase below \SI{3}{mHz}, mainly due to the USO instability and proper time modulations at 1/rev and 2/rev frequencies, while the increase above \SI{3}{mHz} is likely caused by limitations of precise orbit determination and clock solution precision, i.e.~the measurement precision of \epstime.

\begin{figure}
    \centering
    \includegraphics[width=10.7cm]{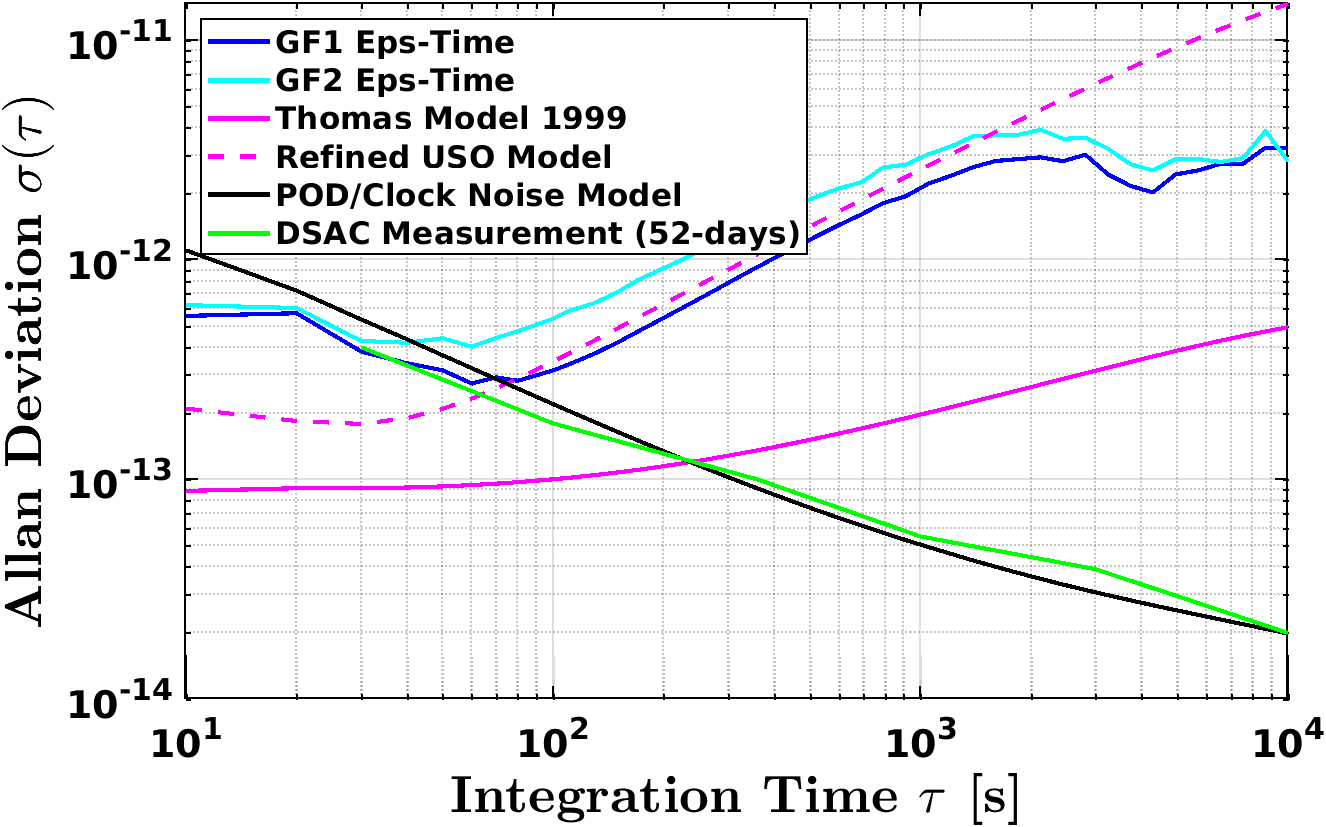}
    \caption{Allan deviations of the most of the traces shown in the right plot of fig.~\ref{fig:propertime}. The additional green trace shows the in-orbit stability of the trapped-ion atomic clock \cite{burt2021demonstration}, which is dominated by GPS measurement noise for the shown values of $\tau$.  }
    \label{fig:AllanDeviation}
\end{figure}

The analytical formula for the USO stability from \cite[B-1]{thomas1999analysis}, shown in solid magenta in fig.~\ref{fig:propertime}, is in poor agreement with the GRACE-FO in-flight data, justifying a refined USO model that we derive ad-hoc as
\begin{align}
    \textrm{ASD}[\delta t_\textrm{USO,mod}](f) = \SI{1}{s} \cdot \sqrt{ 2.556 \cdot 10^{-33}\,\SI{}{Hz^4}/f^5 + 3.325 \cdot 10^{-26}\,\SI{}{Hz} / f^2 }. \label{eq:TimeTagErrorUSO}
\end{align}
The first time-derivative of this USO model is shown by the magenta dashed trace in the right plot of fig.~\ref{fig:propertime}. It is only a rough model which only approximately describes the variability in \epstime~in the spectral domain, because the spectra differ slightly between satellites and the \epstime~ spectra change slightly from day to day (not shown).  Furthermore, we take an educated guess for the measurement precision for eps-time due to POD/Clock errors of the GPS measurement system. The model reads
\begin{align}
    \textrm{ASD}[\delta t_\epsilon](f) = \frac{\SI{1.5}{mm}}{c_0 \cdot \sqrt{f}} + \frac{10\,\textrm{\textmu m} \cdot \textrm{Hz}}{c_0 \cdot \sqrt{f^{2.5}}}, 
    \label{eq:TimeTagErrorPOD2}
\end{align}
and is shown as a black trace in the spectral domain in fig.~\ref{fig:propertime} and as an Allan deviation in fig.~\ref{fig:AllanDeviation}. The model has a rms of $\SI{11}{mm}/c_0$ in the band $\SI{0.1}{mHz}-\SI{0.05}{Hz}$ and (roughly) agrees with the GPS POD/clock measurement noise in the Deep Space Atomic Clock (DSAC) experiment conducted in a low-Earth orbit \cite{burt2021demonstration} with respect to Allan deviations. The noise level of DSAC is consistent with the GRACE-FO USO Allan deviations in fig.~\ref{fig:AllanDeviation} at lowest integration times (i.e. highest frequencies), though it bears mentioning that the GRACE-FO traces are for a single day and exhibit some day-to-day variability (not shown).

The noise models for the timetag error in eq.~(\ref{eq:TimeTagErrorUSO}) and (\ref{eq:TimeTagErrorPOD2}) can produce an error in the phase measurement of KBR and LRI. In order to assess its magnitude, we replace the argument of the phase measurement from correct time $t$ with an erroneous time $t+ \delta t$:
\begin{align}
    \varphi(t) \longrightarrow \varphi(t + \delta t) \approx \varphi(t) + \dot{\varphi} \cdot \delta t.
\end{align}
The effect of the timetag error on the LRI ranging phase can be derived from eq.~(\ref{eq:varphiTWR}), whereas its effect on the LRI range can be determined from eq.~(\ref{eq:RhoNaive}) as
\begin{align}
    \textrm{ASD}[\delta \varphi_\textrm{TWR}] &= \textrm{ASD}[\dot{\varphi}_\xT \cdot \delta t_\xT - \dot{\varphi}_\xR \cdot \delta t_\xR] \approx \dot{\varphi} \cdot \sqrt{2} \cdot \textrm{ASD}[\delta t]  \approx \SI{10}{MHz} \cdot \sqrt{2} \cdot  \textrm{ASD}[\delta t]\\
    \textrm{ASD}[\delta \rho_\textrm{TWR}] &\approx \textrm{ASD}[\varphi_\textrm{TWR} \cdot \lambda_\xR/2] \approx \SI{5}{m/s} \cdot \sqrt{2} \cdot \textrm{ASD}[\delta t], \label{eq:TimetagErrorTWR}
\end{align}
using the fact that the LRI phases measured on transponder and reference satellites have a common and dominant slope of \SI{10}{MHz} (cf.~eq.~(\ref{eq:TransponderPhaseBasic})), and we assumed that the timetag noise on both spacecraft is uncorrelated and has the same stochastic characteristics, such that the differential combination yields $\sqrt{2} \delta t$ in the spectral domain.

The same approach for the KBR at K- and Ka-band yields a timetag induced noise of
\begin{align}
    \textrm{ASD}[\delta \varphi_\textrm{DOWR}^\textrm{K/Ka}] &= \textrm{ASD}[\dot{\varphi}_\xB^\textrm{K/Ka} \cdot \delta t_\xB - \dot{\varphi}_\xA^\textrm{K/Ka} \cdot \delta t_\xA] \\ &\approx \dot{\varphi} \cdot \sqrt{2} \cdot \textrm{ASD}[\delta t]  \approx \sqrt{2} \cdot  \textrm{ASD}[\delta t] \cdot \begin{cases}
\SI{500}{kHz}, &\text{\quad for K-band}\\
\SI{670}{kHz}, &\text{\quad for Ka-band}
\end{cases}\\
    \textrm{ASD}[\delta \rho_\textrm{DOWR}^\textrm{K/Ka}] &\approx \textrm{ASD}[\varphi_\textrm{DOWR}^\textrm{K/Ka} \cdot c_0 / (\nu_\xA^\textrm{K/Ka}+\nu_\xB^\textrm{K/Ka})] \approx \SI{3071}{m/s} \cdot \sqrt{2} \cdot \textrm{ASD}[\delta t].
\end{align}
for the phase $\varphi$ and range $\rho$. Though the timetag error is different for the K- and Ka-phase due to the difference in beatnote frequencies (\SI{500}{kHz} vs. \SI{670}{kHz}), it is the same in terms of range.
When forming the final ionosphere-free DOWR combination, one has to consider that the timetag error is correlated between the K- and Ka-band, such that both contributors add linearly as 
\begin{align}
     \textrm{ASD}[\delta \rho_\textrm{DOWR}] = a^\textrm{K} \cdot \textrm{ASD}[\delta \rho_\textrm{DOWR}^\textrm{K}] + a^\textrm{Ka} \cdot \textrm{ASD}[\delta \rho_\textrm{DOWR}^\textrm{Ka}] \approx \SI{3071}{m/s} \cdot \sqrt{2} \cdot \textrm{ASD}[\delta t], \label{eq:TimetagErrorDOWR}
\end{align}
with the coefficients $a^\textrm{K/Ka}$ given in table~\ref{tab:MWIcoeff}.

Eq.~(\ref{eq:TimetagErrorTWR}) and (\ref{eq:TimetagErrorDOWR}) depend on $\delta t$ and have been used to produce the black timetag curves of LRI and KBR, respectively, in fig.~\ref{fig:NoiseComp}. Traces labeled with \emph{timetag raw} are based on $\delta t=\delta t_\textrm{USO,mod}$ given in  eq.~(\ref{eq:TimeTagErrorUSO}), while \emph{timetag corrected} use the $\delta t=\delta t_\epsilon$ noise from eq.~(\ref{eq:TimeTagErrorPOD2}). As apparent from fig.~\ref{fig:NoiseComp}, the timetag noise in KBR likely is a significant contributor to the increase of KBR-LRI residuals below \SI{10}{mHz}.

However, we note that the timetag error depends on the differential clock jitter between both satellites and our assumption of uncorrelated noise between satellites ($\textrm{ASD}[\delta t_\epsilon] \cdot \sqrt{2}$) neglects potential common-mode rejection of errors that might appear between close-by GPS receiver. Moreover, for the sake of simplicity, we considered just a stochastic USO noise model and omitted potential sinusoidal variations due to the proper time and secular drifts due to uncertainties in the USO frequency for the calculation of the trace labeled \emph{timetag raw}. The proper time effect is not severe, because the differential proper time ($\tau_\xA-\tau_\xB$) between satellites is much smaller than the individual contribution ($\tau_\xA$ or $\tau_\xB$). We regard these approximations as acceptable, as the \emph{timetag raw} trace is anyway rather hypothetical due to the general availability of GPS data to correct the raw values.

\subsection{Carrier Frequency Variations}
\label{sec:CarrierFreqVar}
Carrier frequencies are the laser frequency $\nu_\xR$ in LRI and the K- and Ka-band frequencies $\nu_\textrm{A/B}^\textrm{K/Ka}$ in KBR. Variations in these frequencies correspond to changes in the conversion factor from phase to range, as discussed in sec.~\ref{sec:RangingPhaseLRI}. If the frequency variations are measured and known, they can be accounted for in the conversion and do not falsify the range measurement, when employing eq.~(\ref{eq:Rho2Corr}). For the application in GRACE-FO, the approximation formulas eq.~(\ref{eq:Rho1Corr1}), (\ref{eq:Rho1Corr1approx}) and (\ref{eq:Rho2approx}) usually are  sufficiently accurate.

For the LRI, we assume the following model for the laser frequency at the reference satellite and for a time-series of one day
\begin{align}
    \nu_\xR(t) \approx \langle \nu_\xR(t) \rangle + \langle \nu_\xR(t) \rangle \cdot \left(\frac{\dd \tau_\xR(t)}{\dd t} - 1\right) + \delta \nu_\textrm{cav}(t),  \label{eq:NuRFlightModel}
\end{align}
where $\langle \nu_\xR(t) \rangle$ is the daily mean frequency determined from correlating LRI phase to KBR range (cf.~sec.~\ref{sec:KBRLRIresid}), the second term contains the modulation of the laser frequency due to the proper time $\tau_\xR$ (cf.~eq.~(\ref{eq:NuInGps})) and the last term accounts for the instability of the cavity, which was modelled in sec.~\ref{sec:PhaseObservable}  as $\delta \nu_\textrm{cav}(t) = \dd \tau_\textrm{FV}(t) / \dd t \cdot \langle \nu_\xR(t) \rangle$. The daily mean frequency $\langle \nu_\xR(t) \rangle$ is used as the conversion factor from phase to range (cf.~eq.~(\ref{eq:RhoNaive})) in the LRI1Bv04 processing. The error arising from neglecting the intraday time-variability of the frequency represents a scale factor error as  discussed in sec.~\ref{sec:ScaleError}. Such errors couple into the range to first order as the product of fractional frequency variation and inter-satellite distance $L(t)$. The ranging error due to  $\delta \nu_\textrm{cav}$ is usually called laser frequency noise in the LRI and caused by the limited stability of the cavity resonance frequency, e.g. due to thermal fluctuations. It can be written as
\begin{align}
\quad \textrm{ASD}[\delta\rho_\textrm{FV,corr,LRI}](f) =  \frac{\textrm{ASD}[\delta \nu_\textrm{cav}](f)}{\langle \nu_\xR(t) \rangle}  \cdot L(t)  \quad\textrm{with}  \quad  \textrm{ASD}[\delta \nu_\textrm{cav}](f) = \frac{10^{-15}}{\sqrt{f}} \cdot \langle \nu_\xR(t) \rangle,
\end{align}
which is based on an extrapolation of the LRI ranging noise from high to low frequencies \cite{SPERO2021} and wherein $L \approx \SI{220}{km} \pm \SI{50}{km}$ is the satellite separation. This noise level is in good agreement with pre-flight measurements of the cavity stability \cite{abich2019orbit}. $\textrm{ASD}[\delta\rho_\textrm{FV,corr,LRI}]$ is shown as a solid blue trace in the left plots of fig.~\ref{fig:NoiseComp}. It is labeled as \emph{frequency variations corrected} because this trace assumes that the second proper time term in eq.~(\ref{eq:NuRFlightModel}) was removed. This removal can be accomplished using
\begin{align}
    \rho_\textrm{LRI1Bv04}^\textrm{corr}(t) = \rho_\textrm{LRI1Bv04}(t) + \left(\frac{\dd \tau_\xR(t)}{\dd t} - 1\right) \cdot L(t). \label{eq:LriFreqCorr}
\end{align}
The proper time of the reference satellite $\tau_\xR$ and absolute distance $L$ can be computed from GRACE-FO orbit data (GNI1B data product). The correction is added to the biased range of LRI1B in order to obtain a corrected biased range, which accounts for the variations of  proper time in the reference laser frequency. The spectrum of the raw laser frequency noise can be approximated as the sum of the proper-time term and the cavity stability as
\begin{align}
    \textrm{PSD}[\delta \rho_\textrm{FV,raw,LRI}](f) \approx  \textrm{PSD}\left[ \frac{\dd \tau_\xR}{\dd t} \right] \cdot L^2 + \textrm{PSD}[\delta \rho_\textrm{FV,corr,LRI}],
\end{align}
which is shown as a dashed blue trace in the left plots of fig.~\ref{fig:NoiseComp}.

For the KBR, the carrier frequency variations at low frequencies can be computed directly from \epstime~in the CLK1B data product. Currently, the KBR1B data processing uses daily mean values of the carrier frequencies, which neglects intraday variations. In order to assess the ranging error, we use the USO model introduced in eq.~(\ref{eq:TimeTagErrorUSO}) and shown as a dashed magenta trace in the right plot of fig.~\ref{fig:propertime}, and then compute the contribution as
\begin{align}
\textrm{PSD}[\delta \rho_\textrm{FV,raw,KBR}] = \frac{ \textrm{PSD}[\delta t_\textrm{USO,mod}](f) \cdot (2 \pi f)^2}{2 } \cdot L^2 + \textrm{PSD}\left[ \frac{\dd (\tau_\xA+\tau_\xB)}{2\dd t} \right] \cdot L^2,  
\end{align}
which is depicted as a dashed blue line in the right plot of fig.~\ref{fig:NoiseComp}. The first term on the right side
is again is the product of fractional frequency variations and inter-satellite separation $L$. The factor of 2 in the denominator (cf.~\cite[B-7]{thomas1999analysis}) arises from the fact that two independent USOs are used in the KBR and we assume that both USOs exhibit the same amount of uncorrelated variations. The second summand accounts for the effect of the proper time.

These frequency variations $\textrm{ASD}[\delta \rho_\textrm{FV,raw,KBR}]$ can be corrected by using time-dependent carrier frequencies as defined in table~\ref{tab:MWIcoeff}, which are related to the CLK1B \epstime~by eq.~(\ref{eq:EpsTimeDef}), i.e.
\begin{align}
    \nu_\textrm{A/B}^\textrm{K}(t) = \hat f_\textrm{A/B,USO} \cdot 5076 \cdot \left( 1 - \frac{\dd \epsilon_\textrm{time,A/B}}{\dd t} \right) \\
    \nu_\textrm{A/B}^\textrm{Ka}(t) = \hat f_\textrm{A/B,USO} \cdot 6768 \cdot \left( 1 - \frac{\dd \epsilon_\textrm{time,A/B}}{\dd t} \right)
\end{align}
The time-dependent carrier frequencies should be low-pass filtered in order to remove high frequency fluctuations above \SI{3}{mHz}, which limit the measurement precision of \epstime~and do not stem from real variations in the carrier frequency (cf.~right plot in~fig.~\ref{fig:propertime}). 

The corrected KBR range should ideally be computed in a modified level1A to level1B processing scheme using eq.~(\ref{eq:DOWR2}), which assumes that the re-scaling from phase to range is performed with time-dependent carrier frequencies and also takes the coupling of the phase-bias ($\Delta t_\textrm{FV,DOWR}$) given in eq.~(\ref{eq:DOWR2formulaFV}) into account.

\begin{figure}
    \centering
    \includegraphics[width=10.5cm]{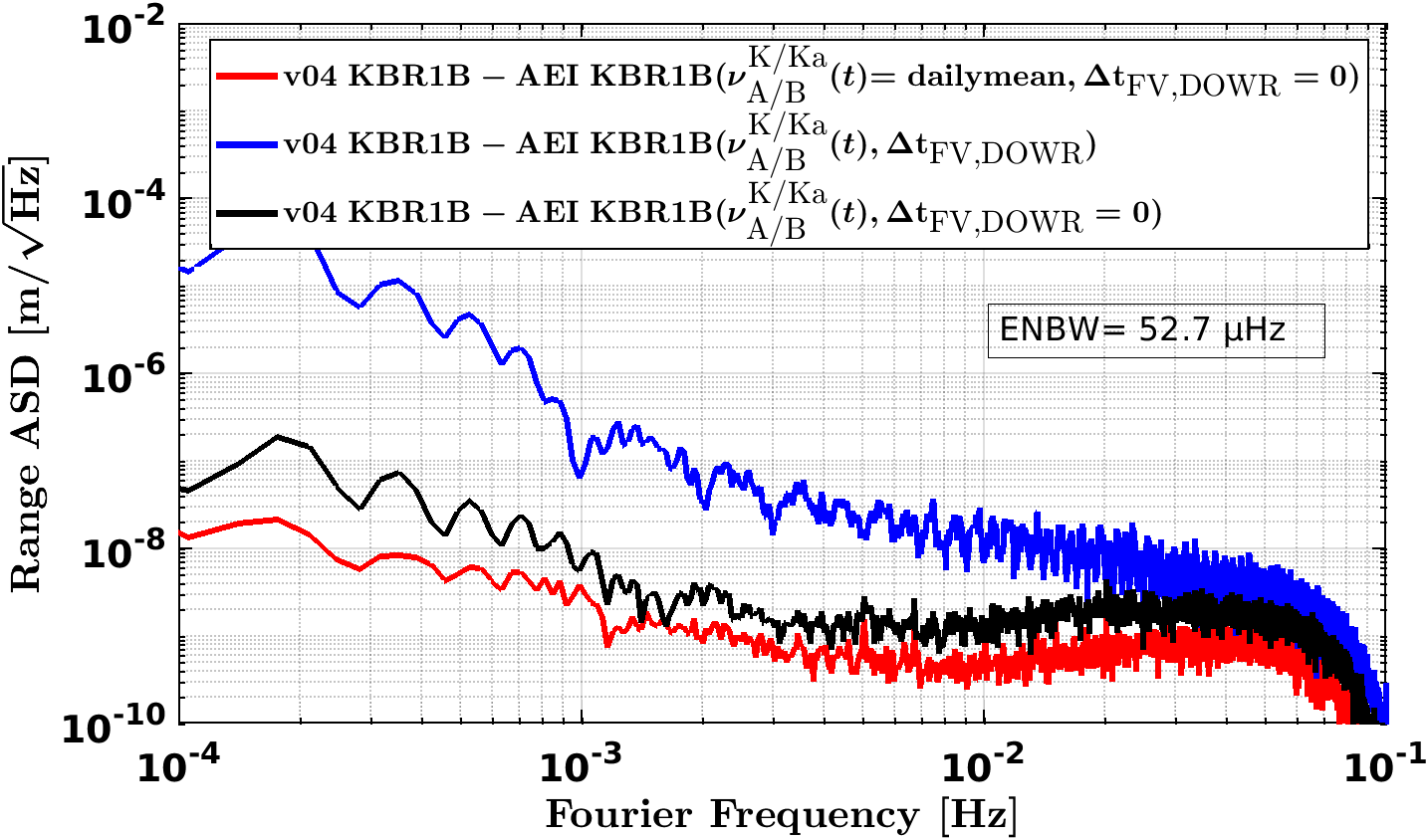} 
    \caption{Difference between the official KBR1B biased range (v04) and a corrected KBR1B data product, from which USO frequency variations were removed. The blue curve shows the difference after frequency variations have been fully removed, while the black trace shows the difference calculated if only the division of the phase with a time-dependent frequency is considered without $\Delta t_\textrm{FV,DOWR}$-term. The red trace serves as a sanity check to show that our self-derived KBR1B product can reproduce the official KBR1Bv04 results up to some negligible deviations, which probably stem from different interpolation methods. The spectrum was computed with GRACE-FO data from 2019-01-01.}
    \label{fig:KbrComparison}
\end{figure}

However, we point out that a direct correction of the KBR1Bv04 range data 
\begin{align}
    \rho_\textrm{KBR1Bv04}^\textrm{corr}(t) \approx \rho_\textrm{KBR1Bv04}(t) + c_0 \cdot \Delta t_\textrm{FV,DOWR}(t), \label{eq:CorrectedKBR1Brange}
\end{align}
might be performed as well, as the magnitude of $c_0 \Delta t_\textrm{FV,DOWR}$ has a much larger effect than dividing the phase by a time-dependent frequency compared to division by a daily mean frequency. This is illustrated in fig.~\ref{fig:KbrComparison}, which also confirms that the difference between both correction approaches is below the KBR noise level ($\approx$ 0.6\,\textmu m/$\sqrt{\textrm{Hz}}$, black trace).

In order to derive the residual effect of USO frequency fluctuations in the KBR ranging data, i.e.~after applying the correction using the modified level1a to level1b conversion, we use the model for the precision of CLK1B \epstime~data derived in sec.~\ref{sec:TimetagError} as $\textrm{ASD}[\delta t_\epsilon]$. The solid blue trace shows the residual KBR frequency variations in the right plot of fig.~\ref{fig:NoiseComp} based on
\begin{align}
\textrm{ASD}[\delta \rho_\textrm{FV,corr,KBR}] = \frac{ \textrm{ASD}[\delta t_\epsilon](f) \cdot 2 \pi f }{\sqrt{2}} \cdot L.
\end{align}
We can conclude that correcting for KBR frequency variations might slightly reduce the KBR-LRI residuals at frequencies between \SI{0.1}{mHz} and \SI{1}{mHz} (dashed blue vs solid blue trace). However, the change is expected to be rather small as the timetag error (solid black trace) becomes limiting.

\section{Impact of Frequency Variations on KBR-LRI residuals}
\begin{figure}
    \centering
    \includegraphics[width=14.0cm]{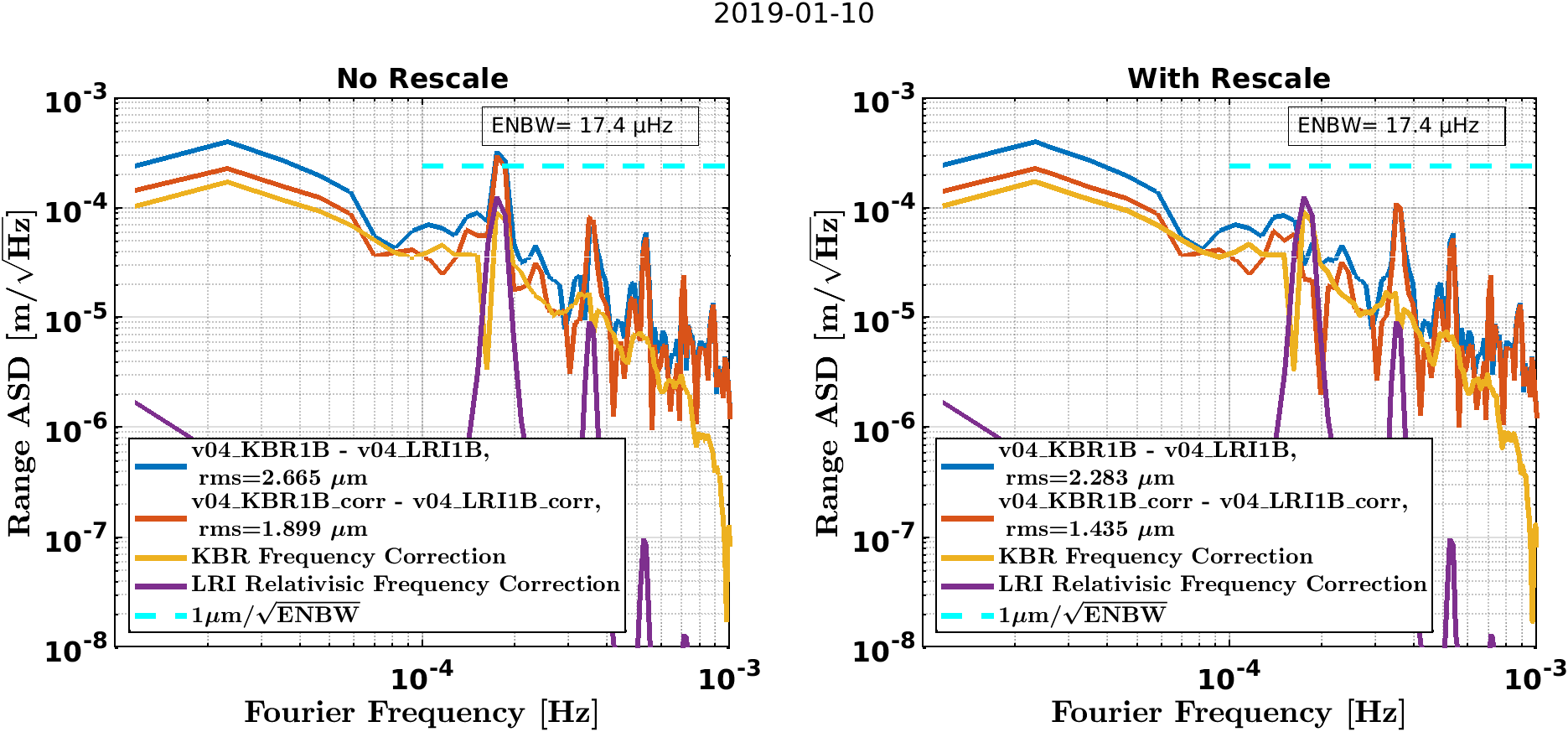} \\
    \includegraphics[width=14.0cm]{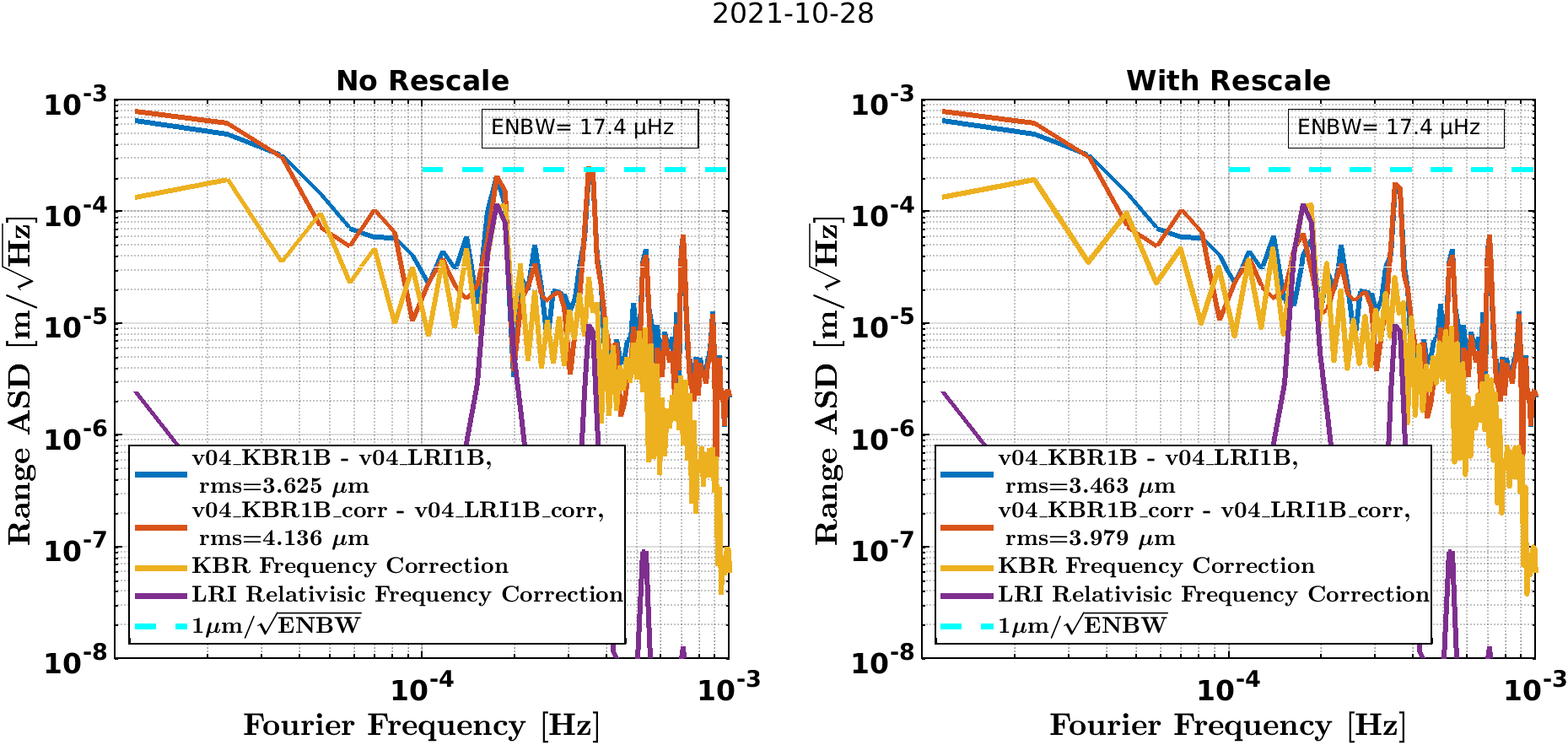}
    \caption{KBR-LRI residuals at low frequencies without applying the frequency corrections (blue) and with applied LRI and KBR frequency correction (red). The corrections are shown as yellow and purple traces. }
    \label{fig:CorrectionImpactAsd}
\end{figure}

We have corrected the relativistic (proper-time) frequency variations in LRI and the KBR frequency variations as discussed in previous sub-section~\ref{sec:CarrierFreqVar} on a daily basis for most of the days between December 2018 and March 2022 in the GRACE-FO v04 data. Days where LRI1B, KBR1B or CLK1B data was incomplete, e.g. due to gaps from instrument reboots or diagnostic events, were skipped. For each day, we evaluated the spectrum of KBR-LRI residuals as shown exemplarily in fig.~\ref{fig:CorrectionImpactAsd}. On day 2019-01-10, the spectrum of KBR-LRI residuals was reduced when applying the correction, while day 2021-10-28 shows an increase in the residual level (compare red and blue traces).

As a metric, we use the root-mean-square (rms) value of KBR-LRI residuals at low frequencies, which is obtained by integrating the power spectral density (squared ASD) for frequencies below \SI{1}{mHz}. The plots titled \emph{With Rescale} (right column) indicate that we re-estimated the differential scale and time-shift between LRI and KBR together with a linear trend. This is in particular necessary for days before 2019-08-12, as the LRI1B data is sometimes not properly scaled and time-shifted in this period.

\begin{figure}
         \centering
    \includegraphics[width=13.5cm]{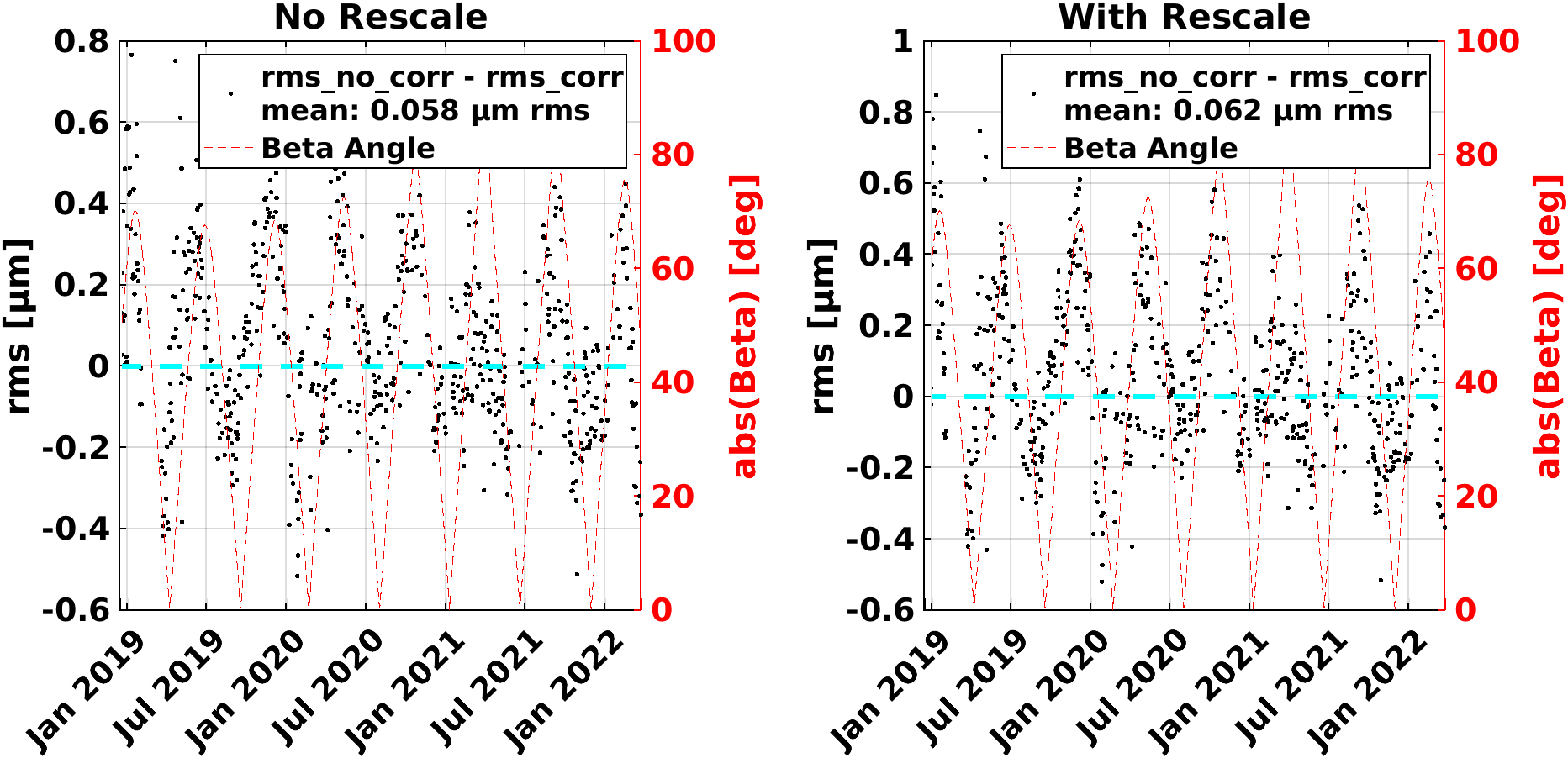}
    \caption{Daily change in the low frequency rms value ($<\SI{1}{mHz}$) of the KBR-LRI residuals when applying the KBR and LRI frequency variation correction. Both plots show the difference in rms value before and after the correction.
    The left plot uses original LRI1Bv04 data without re-estimating the scale and time-shift between LRI and KBR, while the right plot assumes re-estimation (rescale) of scale, time-shift and linear trend on a daily basis. Values above the cyan dashed line indicate a reduction in rms value due to the correction. On days with a high absolute value of beta angle, i.e.~close to or in full-sun phase of the satellites, the frequency correction yields the largest rms reductions, e.g.~in January 2019.}
    \label{fig:CorrectionImpact}
\end{figure}

\begin{figure}
    \centering
    \includegraphics[width=13.5cm]{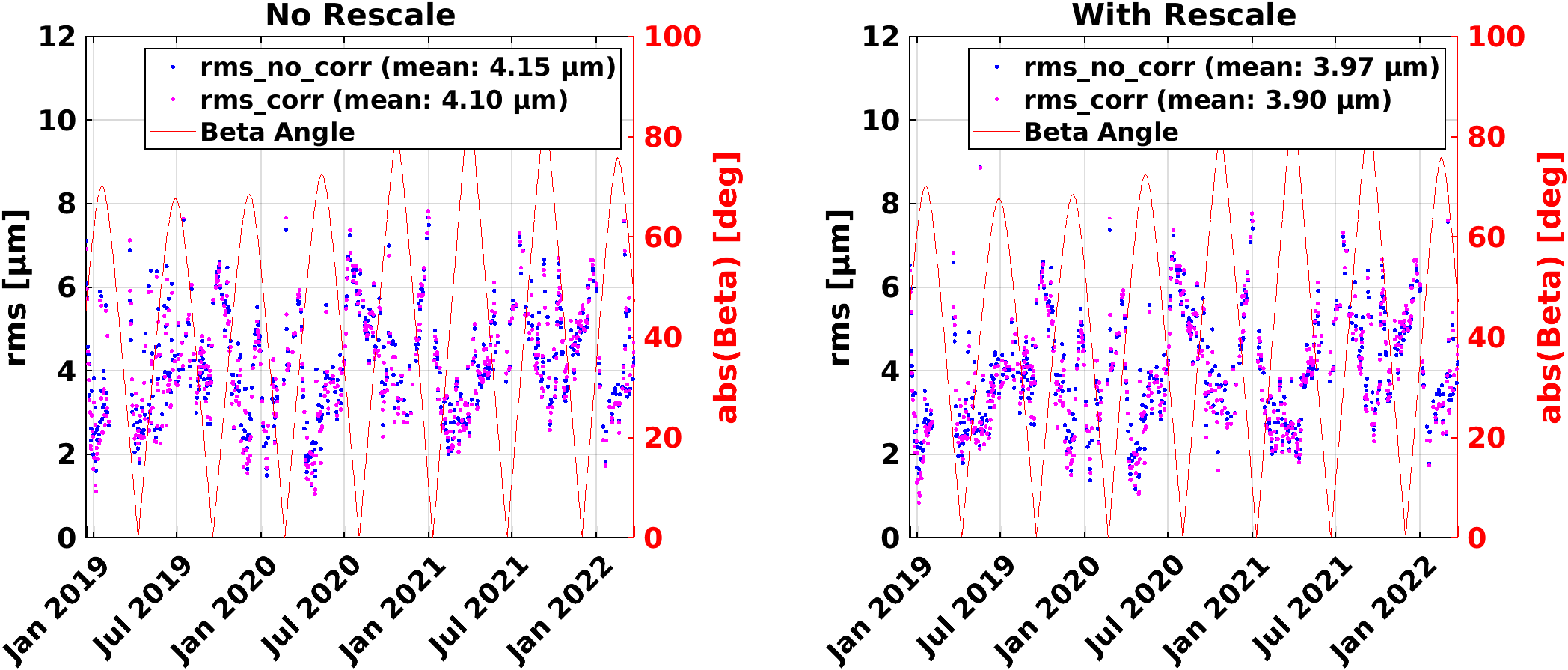}
    \caption{Daily absolute value of low frequency ($<\SI{1}{mHz}$) rms of the KBR-LRI residuals. The rms value of residuals is low when the beta angle is large, i.e.~when satellites are close to or in full-sun phase and not entering shadow. The left plot uses original LRI1Bv04 data with its scale and time-shift, while the right plot assumes re-estimation (rescale) of scale, time-shift and linear trend on a daily basis.}
    \label{fig:CorrectionImpactTotalRms}
\end{figure}

The change in daily rms value of LRI-LBR residuals at the low frequency due to LRI and KBR frequency correction is shown in fig.~\ref{fig:CorrectionImpact}. For January 2019, the rms value could be lowered noticeably, but, the values oscillate with the beta angle of the orbit, i.e.~the angle between orbital plane and Sun direction. The absolute rms values are depicted in fig.~\ref{fig:CorrectionImpactTotalRms} and exhibit the oscillatory pattern as well. Both plots indicate only a marginal reduction in rms due to the frequency correction, e.g.~a change in mean rms from 3.97\,\textmu m to 3.90\,\textmu m.

We repeated the analysis with more empirical parameters being estimated in the daily rescale step, namely, the 1/rev, 2/rev and 2/day oscillation amplitudes in terms of sine and cosine components and a quadratic trend, in addition to the differential scale, time-shift and linear trend already in use. The 1/rev and 2/rev frequencies in the ranging data are likely to have significant errors, as environmental parameters such as temperature, inertial orientation and magnetic field, predominately vary at these frequencies. The 2/day parameter was added due to the CLK1B \epstime~exhibiting a pronounced 2/day feature on several days (cf.~fig.~\ref{fig:propertime}) which could be related to Earth's rotation period or to the orbit period of satellites in the GPS constellation.

\begin{figure}
    \centering
    \includegraphics[width=13.5cm]{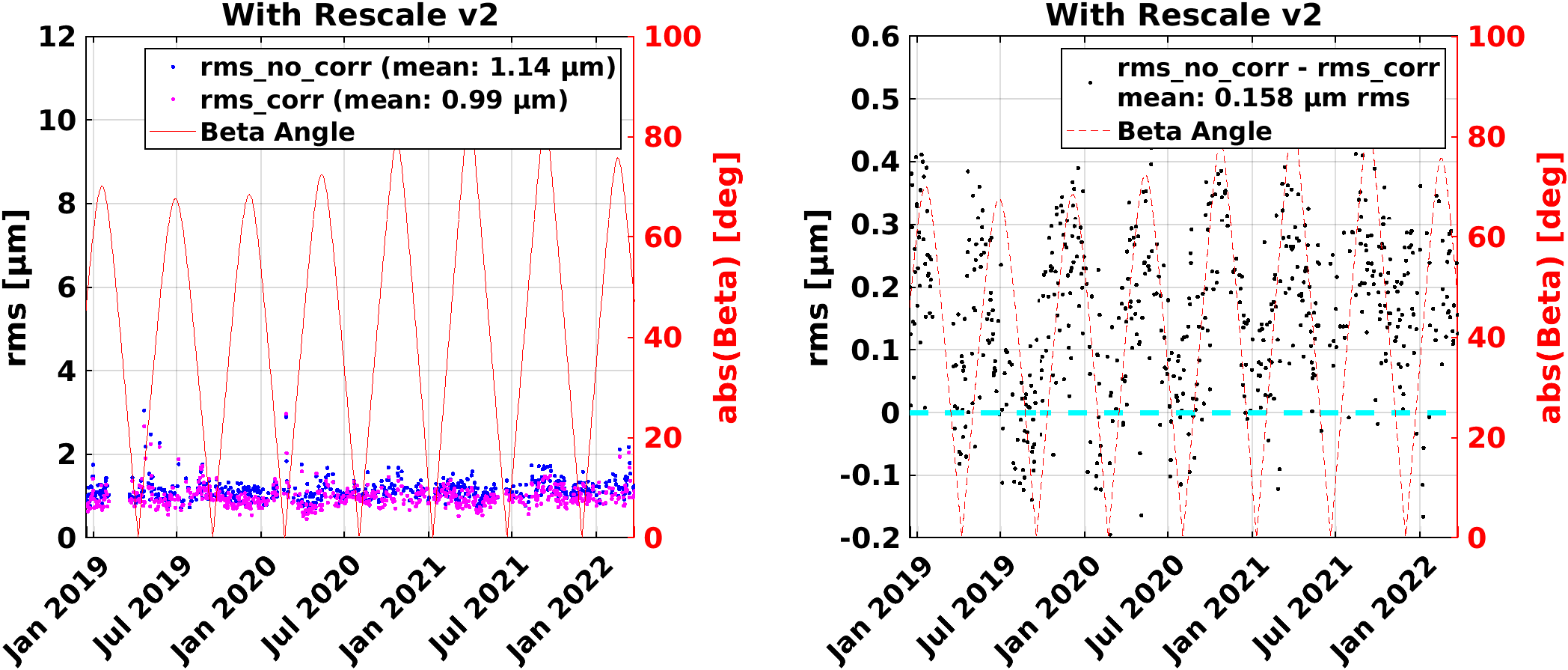}
    \caption{Results as in fig.~\ref{fig:CorrectionImpact} and \ref{fig:CorrectionImpactTotalRms}, but now using more empirical parameters for the daily estimates: 1/rev, 2/rev, 2/day, scale, time-shift and with linear and quadratic trends.}
    \label{fig:CorrectionRefineds}
\end{figure}

We then estimated the KBR-LRI residuals together with the many empirical parameters again, once without the frequency correction and once with the correction for individual days. The results are shown in fig.~\ref{fig:CorrectionRefineds}. The additional empirical parameters absorbed the oscillatory beta angle-related behavior in the rms value to a large extent and lowered the rms value from approx.~4\,\textmu m to approx.~1\,\textmu m, as had been expected.
A change in rms value due to the frequency correction thereby became visible more clearly, i.e.~from mean rms of 1.14\,\textmu m to 0.99\,\textmu m (cf.~legend in figure). This 0.15\,\textmu m change in rms in the 1\,mHz bandwidth, which would imply a spectral density improvement of 0.15\,\textmu m/$\sqrt{\textrm{1\,mHz}}=$\,4.7\,\textmu m/$\sqrt{\textrm{Hz}}$ assuming white noise behavior, excludes effects at 1/rev, 2/rev and 2/day, because KBR-LRI residuals at these frequencies are highly reduced due the empirical parameters. Therefore, this change should  be understood as a lower bound as it neglects the frequencies where the correction has most of its signal.

The reduction in rms demonstrates that applying the frequency corrections, i.e.~considering intraday variability of the carrier frequencies, slightly improves the agreement of LRI and KBR ranging data.
We also emphasize that the relativistic effect due to the proper time, i.e. 0.9\,\textmu m-peak at 1/rev (cf.~eq.~(\ref{eq:LriProperTimeError})), is omitted in both instruments, if no frequency correction is applied. If the corresponding frequency correction is applied to both instruments, the 0.9\,\textmu m-peak is present in both instruments. Thus, it never appears in the KBR-LRI residuals. However, neglecting the effect would introduce an error compared to a true error-free range.

The effect of the frequency corrections on LRI and KBR is small, i.e.~at the micron scale at very low frequencies, which means it should have negligible impact on gravity fields, as typical pre-fit and post-fit residuals of gravity fields are much larger than the KBR-LRI residuals. However, these corrections reduce the errors in the instruments and are a necessary step towards a better understanding of the KBR-LRI residuals at low frequencies, e.g.~when studying the tone errors of instruments at 1/rev and 2/rev frequencies.

\section{Summary and Conclusions}
GRACE Follow-On offers the unique opportunity to compare two almost independent ranging instruments, the KBR and the LRI. Both instruments rely on readout of phase changes present in the beat note of interfering microwave or near-infrared laser radiation. The measured phase values can be converted to a range with units of meter using the respective carrier frequencies $\nu_\textrm{A/B}^\textrm{K/Ka}$ or $\nu_\xR$.

In this paper, we have derived and analyzed different formulas for converting the phase to a range when the time-dependency of the carrier frequency is taken into account. It turned out that the previously suggested approach of differentiating the phase, rescaling the derivative using the frequency and performing an integration afterwards is not exact, since a summand was neglected. In this work, we presented the exact formula, which also is likely to be relevant for future laser ranging systems. Besides the analytically exact formula, several approximate formulas can be used that yield results with picometer precision, which is usually sufficient in a GRACE-like context.

We stressed that the time-variability of the carrier frequency $\nu$, if it is neglected in the phase-to-range conversion, yields an error that is proportional to the satellite separation $L$, commonly called laser frequency noise in the LRI. The time-variability is, on the one hand, caused by relativistic effects from the proper time, which is common to both LRI and KBR, and on the other hand by the instability of the optical cavity for LRI and by the instability of the USO for KBR. In the case of the KBR, the microwave carrier frequencies can be determined in post-processing from GPS precise orbit and clock error determination.

In the second part of this paper, the difference between measured range of LRI and KBR in the official v04 data products was compared to some potential noise and error contributors. It was pointed out that the KBR-LRI residuals are bound at frequencies above a few mHz to a level of approx.~0.6\,\textmu m/$\sqrt{\textrm{Hz}}$ due to the CNR-limited phase readout noise from KBR, though the analytical model predicts a noise level of approx.~0.4\,\textmu m/$\sqrt{\textrm{Hz}}$. This small discrepancy has not been understood so far and is caused by the K-band observations, while Ka-band measurements seem to be consistent with the model.

According to our analysis, two effects are limiting the KBR-LRI residuals at low frequencies. Residual timetag errors in the KBR system after applying the GPS-derived clock correction, which cannot be further improved with post-processing, and the intraday carrier frequency variations in the KBR and LRI, which can be corrected using the formulas introduced in the first half of this paper. The LRI frequency variations can only be corrected for the proper time effect using GNI1B orbit data, which contain dominant 1/rev and 2/rev components. In GRACE-FO, there is no direct measurement of the absolute laser frequency, and thus, the laser frequency is assumed to be static on a daily basis and its daily values are derived from correlating LRI and KBR observations. For KBR, the microwave frequencies are derived from the USO and potential frequency variations can be determined with respect to GPS data. The corresponding KBR frequency correction accounting for intraday variability can be computed directly from the CLK1B data product and exhibits a continuous spectrum as well as a 1/rev peak due to the proper time effect.

In the end, we applied the KBR and LRI frequency corrections and were able to demonstrate a slight improvement in the agreement between KBR and LRI range, i.e.~the rms value of KBR-LRI residuals below 1\,mHz was reduced by 0.15\,\textmu m. However, that figure excludes the 1/rev, 2/rev and 2/day frequencies as well as linear and quadratic drifts. Hence, the actual reduction might be larger. 

Moreover, applying the frequency correction to KBR and LRI should reduce the error at the 1/rev frequency by approx.~0.9\,\textmu m-peak due to the relativistic proper time effect, as the current regular processing that generates v04 data is not accounting for this effect in LRI and KBR data.

We expect that the micron-scale effects at low frequencies discussed here have negligible impact on the current gravity field determination, as typical pre-fit and post-fit residuals of gravity field determination are much larger than direct KBR-LRI residuals. However, the corrections are of interest, if one attempts to understand the low-frequency behavior of the instruments, e.g. the tone errors at 1/rev and 2/rev frequency. These corrections might also be important for future missions employing laser ranging that aim for higher precision at sub-mHz frequencies.

\vspace{6pt} 



\authorcontributions{Conceptualization, Vitali Müller; Funding acquisition, Gerhard Heinzel; Investigation, Markus Hauk, Malte Misfeldt, Laura Müller and Henry Wegener; Project administration, Gerhard Heinzel; Writing – original draft, Vitali Müller and Yihao Yan; Writing – review \& editing, Markus Hauk, Malte Misfeldt, Laura Müller, Henry Wegener and Gerhard Heinzel.}

\dataavailability{Data can be made available upon reasonable request.} 

\funding{This work has been supported by: The Deutsche Forschungsgemeinschaft (DFG, German Research Foundation, Project-ID 434617780, SFB 1464). Clusters of Excellence “QuantumFrontiers: Light and Matter at the Quantum Frontier: Foundations and Applications in Metrology” (EXC-2123, project number: 390837967); the European Space Agency in the framework of Next Generation Geodesy Mission development and ESA's third-party mission support for GRACE-FO; the Chinese Academy of Sciences (CAS) and the Max Planck Society (MPG) in the framework of the LEGACY cooperation on low-frequency gravitational-wave astronomy (M.IF.A.QOP18098).  }

\conflictsofinterest{The authors declare no conflict of interest.}



\appendixtitles{no} 
\appendixstart
\appendix
\section[\appendixname~\thesection]{}
\subsection[\appendixname~\thesubsection]{}
\label{sec:Sim}
In GRACE-like missions, the inter-satellite distance $L(t)$ can be modelled with the following dominant components
\begin{align}
    L(t) = L_0 + L_1 \sin(2 \pi f_\textrm{orb} t) + L_d \cdot t
\end{align}
such that the propagation time can be described as
\begin{align}
    \Delta t(t) \approx L(t)/c_0.
\end{align}

We assume the following frequency model for the LRI laser with constant, oscillatory and linear drift components:
\begin{align}
    \nu_\xR(t) = \nu_0  + \nu_1 \sin(2 \pi f_{\nu 1} t) + \nu_d \cdot t. 
\end{align}
Finally, the phase of the laser can be written as
\begin{align}
\Phi_\xR(t) = \int_{0}^t \nu(t^\prime)~\dd t^\prime = \nu_0 t - \frac{\nu_1 \cos(2 \pi f_{\nu 1} t) }{2 \pi f_{\nu 1} } + \frac{\nu_d \cdot t^2}{2}
\end{align}
 or as
 \begin{align}
     \varphi_\textrm{TWR}(t) = \Phi_\xR(t) - \Phi_\xR(t-\Delta t)
 \end{align}
in terms of the ranging phase $\varphi_\textrm{TWR}$. These analytical models of the range and phase, with the numerical values given in table~\ref{tab:SIMparam}, can be used to validate the expressions derived in sec.~\ref{sec:RangingPhaseLRI}. 
 
\begin{table}[H] 
	\caption{Numerical values for a GRACE-FO scenario. \label{tab:SIMparam} }
\begin{tabularx}{\textwidth}{CCC}
\toprule
\textbf{Name}	& \textbf{Formula/Value}	& \textbf{Comment}\\
\midrule
		$\nu_0$   &  \SI{282}{THz} & LRI optical frequency (1064.5~nm)  \\
		$\nu_1$   &  $4 \cdot 10^{-12} \cdot \nu_0$ & proper time 1/rev variation (cf.~fig.~\ref{fig:propertime})  \\
		$\nu_d$   & $3.6 \cdot 10^{-15} \cdot 1/s \cdot \nu_0 $ & assumption, \SI{87}{kHz/day}  \\
		$L_0$   & $\SI{220}{km}$  & satellite separation DC   \\
		$L_1$   & $\SI{400}{m}$  &  1/rev range variation   \\
		$L_d$   & $\SI{0.01}{m/s}$  &  linear range drift \\ 
		$f_\textrm{orb}$   & $\SI{0.176}{mHz}$  &  orbit frequency, 1/rev \\

\bottomrule
\end{tabularx}
\end{table}

\begin{adjustwidth}{-\extralength}{0cm}

\reftitle{References}


\bibliography{mainKbrLriPaperV1}

%


\end{adjustwidth}
\end{document}